# Multi-contact synapses for stable networks: a spike-timing dependent model of dendritic spine plasticity and turnover

Moritz Deger[1,2], Alexander Seeholzer[1], Wulfram Gerstner[1]

1: School of Computer and Communication Sciences and School of Life Sciences, Brain Mind Institute, École Polytechnique Fédérale de Lausanne, 1015 Lausanne EPFL, Switzerland
2: Institute for Zoology, Faculty of Mathematics and Natural Sciences, University of Cologne, 50674 Cologne, Germany

**Abstract**

Excitatory synaptic connections in the adult neocortex consist of multiple synaptic contacts, almost exclusively formed on dendritic spines. Changes of dendritic spine shape and volume, a correlate of synaptic strength, can be tracked in vivo for weeks. Here, we present a combined model of spike-timing dependent dendritic spine plasticity and turnover that explains the steady state multi-contact configuration of synapses in adult neocortical networks. In this model, many presynaptic neurons compete to make strong synaptic connections onto postsynaptic neurons, while the synaptic contacts comprising each connection cooperate via postsynaptic firing. We demonstrate that the model is consistent with experimentally observed long-term dendritic spine dynamics under steady-state and lesion induced conditions, and show that cooperation of multiple synaptic contacts is crucial for stable, long-term synaptic memories. In simulations of a simplified network of barrel cortex, our plasticity rule reproduces whisker-trimming induced rewiring of thalamo-cortical and recurrent synaptic connectivity on realistic time scales.

Excitatory synapses on pyramidal cells of the mammalian neocortex are almost exclusively made onto postsynaptic dendritic spines (see [1, 2]). Dendritic spines are small protrusions of the dendrite that vary in size and shape, and are subject to ongoing plasticity in both the developing and the adult brain [3, 4, 5, 6, 7, 8]. A central role of dendritic spines may be that the compartment of the spine allows for localized, synapse-specific plasticity mechanisms [1, 2]. Furthermore, plastic dendritic spines may allow neurons to select the subset of inputs from nearby axons of candidate presynaptic cells [9], so as to optimally process stimuli under the constraints of limited total tissue volume available for synaptic connections [10, 11]. Although dendritic spines show considerable volatility and may be formed and eliminated within a couple of days [12], some spines are maintained over long periods of time [12] and may support lifelong memories [13]. During learning, dendritic spines in the neocortex change selectively, and if potentiated spines are selectively shrunk by optogenetic means, learning is disrupted [14].

Time-lapse imaging of changes of spines on time-scales of seconds, hours and days to weeks [15, 12, 16, 4, 5, 17, 18] has shown that spine removal is tightly linked to the dynamics of the spine volume and to synaptic efficacy. Despite the ongoing dendritic spine turnover response patterns of neuronal networks can be long-term stable [19]. While past modeling studies of synaptic consolidation indicate how memories could be stored beyond the first hour after plasticity induction [20, 21, 22, 23], spike-timing dependent plasticity (STDP) based models of synaptic plasticity [24, 25, 26] have not yet been linked quantitatively to dendritic spine plasticity and turnover [5], lifetime of synaptic contacts, and observed statistics of synaptic contact numbers [27, 28].

Synapses between pyramidal neurons in the somatosensory cortex of the rat are sparse and the distribution of the number of synaptic contacts for pairs of pre- and postsynaptic neurons has a characteristic bimodal shape [27, 28, 29]: most presynaptic neurons make no contact at all with a given postsynaptic neuron, but those that do are likely to establish 4 or more contacts [27]. However, the number of potential synaptic contacts between a pair of neurons, based on the proximity of axons and dendrites in reconstructed neuronal morphologies, is always unimodally distributed [28, 30, 29] with a most likely value of 1 and an estimated mean value significantly greater (Fig. 1D) than the mean number of actual contacts observed in experiments. This puzzling observation led to the hypothesis that the observed difference between potential and actual contact numbers cannot arise from independent synapse formation, but instead requires a mechanism of cooperation between synaptic contacts onto the same neuron [28]. We wondered whether the cooperativity between synaptic contacts would necessarily imply a novel biochemical mechanism or whether known principles of STDP are sufficient to explain the bimodal distribution of contact numbers in sparsely connected, stable neocortical networks.

We propose a unifying approach to dendritic spine formation, plasticity and removal, in which STDP of single synaptic contacts provides the crucial cooperative mechanism regulating structural plasticity. Synaptic contacts connecting a neuron A to another neuron B share the same pre- and postsynaptic neuronal activity: presynaptic action potentials (spikes) cause Glutamate release and eventually leave a biochemical trace at each dendritic spine, while information about postsynaptic spikes may reach the same dendritic spines by means of backpropagating action potentials [31, 32].Since in some preparations spine plasticity [33, 34] and spine formation [18] are visible within less than a minute after stimulation, structural spine plasticity cannot be assumed to be slower than STDP, in contrast to previous models [35, 36, 37, 38, 39, 40].

Here, we present a quantitative model that links continuous synaptic plasticity of dendritic spines to discrete structural plasticity of synaptic contact formation and removal. We show that our model reproduces experimentally observed steady-state properties of dendritic spine plasticity and turnover in the somatosensory cortex of adult rodents. While existing theoretical models of the bimodal distribution of synaptic contacts in networks are algorithmic [28, 30], our model suggests that STDP is sufficient to generate the bimodal distribution observed in experiments. Our model predicts that a synaptic contact that is part of a connection consisting of multiple synaptic contacts is more stable than an isolated synaptic contact. Moreover, a recurrent thalamo-cortical network model exhibits long-term stable structure despite ongoing plasticity, and lesion-induced rewiring on time scales of days and weeks.

# Results

## A combined model of STDP and structural plasticity

In a first set of simulated experiments, we model the plasticity of the synapses from $N = 1000$ presynaptic neurons onto a single postsynaptic neuron (Fig. 1A). All presynaptic model neurons fire as independent Poisson processes with a rate of 5/s. A connection from a presynaptic neuron $j$ to the postsynaptic cell may consist of several synaptic contacts $k$, with $1 \leq k \leq n_j$, where $n_j$ is the number of potential contacts [28, 30]. Each of these contacts is described by a unit-less weight $w_{jk}$. The total weight of the connection from a presynaptic neuron $j$ is given by the sum of the weights $w_{jk}$ over all its contacts $k$, $w_j = \sum_{k=1}^{n_j} w_{jk}$. The contact weight $w_{jk}$ in our model describes how much the specific contact $k$ contributes to firing the postsynaptic cell, and can be viewed as representing the dendritic spine volume which is in turn strongly correlated to the AMPA receptor content in the postsynaptic density [5].

Our plasticity model (Fig. 1B) describes the temporal dynamics of the weight $w_{jk}$ of a synaptic contact by the differential equation

$$\frac{d}{dt} w_{jk}(t) \quad = \quad F(S_j,\, S_{\text{post}},\, z_{jk}) - \alpha w_{jk}(t), \tag{1}$$

where $F$ is a functional of the pre- and postsynaptic spike trains $S_j$ and $S_{\text{post}}$, respectively, and of stochastic synaptic transmission $z_{jk}$ at the contact. The parameter $\alpha > 0$ describes a slow decay of the weight. The model (1) is a local, spike-timing dependent plasticity rule that we imagine to be realized by the biophysics of dendritic spines in combination with that of the presynaptic terminal. In our model pre- or postsynaptic spikes each leave an exponentially decaying trace ($r_{jk}$ and $r_{\text{post}}$, respectively) at the synaptic contact (dendritic spine), with a short time constant $\tau$ (Fig. 1C). A third filter trace $C_{jk}$ with a longer time constant $\tau_{\text{slow}}$ tracks the product of $r_{jk} \cdot r_{\text{post}}$ as a measure of the correlation of pre- and postsynaptic spike trains. A fourth filter trace $R_{\text{post}}$ (with time constant $\tau_{\text{slow}}$) tracks the postsynaptic firing rate. The combination of these traces gives rise to the STDP model summarized by $F$ in Eq. (1), see Methods for further details. Mathematical analysis shows that this plasticity rule establishes a target level for both firing rate and spike-timing correlations (see Model analysis in Methods and Supplementary Information).

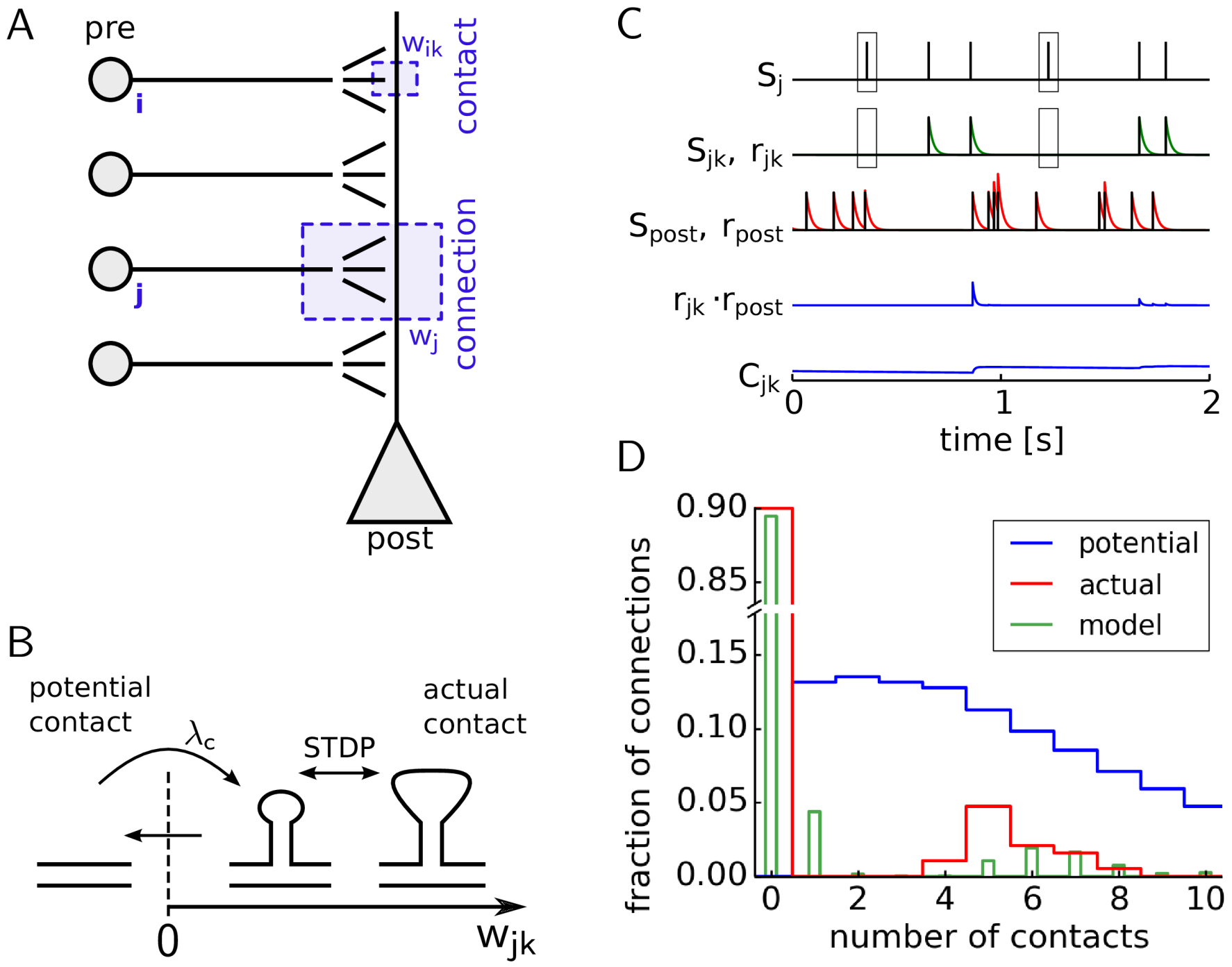

Figure 1: **Model overview. A:** A postsynaptic neuron (post) receives synaptic connections from several excitatory neurons (pre). Each connection consists of multiple synaptic contacts. The synaptic efficacy (weight) of contact $k$ in connection $j$ is denoted as $w_{jk}$. The sum of contact weights $w_{jk}$ is the total weight $w_j$ of this synaptic connection. **B:** Individual contact weights $w_{jk}$ take continuous, positive values, and change in time according to spike-timing dependent plasticity. A small weight $w_{jk}$ corresponds to a dendritic spine with a small volume, or a thin spine, and a large weight $w_{jk}$ corresponds to a large, or mushroom-shaped dendritic spine. If $w_{jk}$ is greater than 0 we call the contact an actual contact. In contrast, if at any time $w_{jk}$ becomes 0, the contact is pruned and its weight is kept fixed at $w_{jk} = 0$. A contact with $w_{jk} = 0$ is called a potential (but inactive) contact. It may be transformed into an actual contact by setting $w_{jk}$ to a positive value at random times, with a rate $\lambda_c$ (creation rate). **C:** Components of the plasticity model (top to bottom): presynaptic spike train $S_j$; transmitted spike train $S_{jk}$ at the contact (black, random synaptic failures occur, indicated by black boxes), and its filtered trace $r_{jk}$ (green); postsynaptic spike train $S_{\text{post}}$ (black) and its filtered trace $r_{\text{post}}$ (red); product (correlation) term $r_{jk} \cdot r_{\text{post}}$ composed of pre- and postsynaptic trace (blue); low-pass filtered trace $C_{jk}$ of the product term $r_{jk} \cdot r_{\text{post}}$. Synaptic contact plasticity is determined by $C_{jk}$. **D:** Reference distributions of the number of actual (red) [27] and potential (blue) [28] synaptic contacts for pairs of neurons in the adult somatosensory cortex (recurrent connections of layer 5 pyramidal neurons, truncated to $n_j \leq 10$ and renormalized). The steady state distribution generated by our model is shown in green (data pooled over 150 days of simulation); the distribution of potential contacts in the model is matched to the blue line.

As a result of ongoing STDP, synaptic model weights can increase or decrease. As soon as a contact weight $w_{jk}(t)$ decreases to zero in the simulation, its dynamics cease and it turns into a potential but inactive contact. However, with a rate of $\lambda_c = 0.019$/day, each potential contact ($w_{jk} = 0$) may randomly be transformed into an active contact. In such an event the weight $w_{jk}$ of the newly created contact is set to a low, non-zero value $w_c$. After creation the contact weight remains at the value $w_c$ for the duration of the "period of grace" [41] of $\tau_{gp} = 15$ min, after which its dynamics again follow (1) (see Fig. 1B).

In our model, the number of potential synaptic contacts $n_j$ from a presynaptic neuron $j$ to the postsynaptic model neuron lies between 1 to 10, and is distributed as estimated for recurrent connections of layer 5 pyramidal neurons in rat somatosensory cortex [28], so that the blue line in Fig. 1D represents the distribution of potential contacts $n_j$ in both model and experiment. The model is insensitive to the exact maximum value of $n_j$, and values of $n_j$ up to 20 [28] would give equivalent results, albeit at increased computational cost. In the following, we describe how our plasticity model turns potential contacts into stable weights through the mechanism of STDP-mediated cooperation.

## Steady state synapse dynamics under STDP-mediated cooperation

After an initial transient at the beginning of the simulation, the synaptic contacts fluctuate around a steady state. Sample trajectories of all contacts from a single presynaptic neuron to the postsynaptic cell are shown in Fig. 2A-B. All initial contacts of this presynaptic neuron remain over the course of 150 days of simulation, occasionally new contacts are formed, which mature or are quickly removed (see Fig. 2B). The firing rate of the postsynaptic neuron, as well as the total synaptic weight and the average number of contacts are tightly controlled by the plasticity rule (Fig. 2C and Methods). The average number of active contacts $\langle N_{\text{contacts}} \rangle$, corresponding to the spine density, is stationary in the model consistent with experiments [42, 12, 43, 5].

Within one day, the relative change of the contact weight ranges from $-100\%$ to $500\%$ (Fig. 2D) but the relative fluctuations decrease with increasing contact weight, consistent with long-term time-lapse imaging data of dendritic spine volume in vitro [4]. We quantify the statistics of contact weight changes by computing the mean and standard deviation of the change within one day, as a function of the weight. Consistent with experimental measurements of dendritic spine volume [4] the mean contact change (Fig. 2E, black) is positive for an intermediate range of contact weights (spine volumes) and becomes negative for large weights (large volumes), and the standard deviation of the changes (Fig. 2E, red) is rather homogeneous for all weights (volumes) but increases slightly for very large weights. The mean change of contacts of very small weight is negative in our model, in contrast to previous experiments [4], but this difference might be due to experimental difficulties of observing very small spines (see Discussion).

As synaptic connections in our model consist of several synaptic contacts, we asked whether there is a systematic relation between synaptic weight and the number of contacts in the steady state (Fig. 2F). Indeed the synaptic weight is strongly correlated with the number of active contacts: Strong synaptic connections consist of at least five synaptic contacts, and connections with less than three contacts are physiologically weak (Fig. 2F, red, left axis). Moreover, individual contacts in connections made of more than 3 active contacts tend to be stronger than those made of less than 3 (Fig. 2F, blue, right axis).

We further assess the statistical properties of the steady state by multiple measures. First, the distribution of active synaptic contact numbers per connection (Fig. 3A) is bimodal, in line with experimental findings [27, 28], cf. Fig. 1D. Second, the turnover ratio of synaptic contacts in the model is $0.176 \pm 0.018$/day (mean $\pm$ STD) consistent with the values found experimentally in somatosensory cortex [12, 5]. Third, a stable distribution of contacts weights (Fig. 3B) is formed.

The probability of a synaptic contact to survive for 8 consecutive days, irrespective of whether the contact is newly created, weak or strong, depends strongly on the number of active contacts in the connection that the contact is part of (Fig. 3C). In other words, contacts within a connection cooperate and stabilize each other. Tracking of individual synaptic contacts that existed at $t = 0$ (Fig. 3D) reveals a time course of the number of surviving synapses that is qualitatively consistent with long-term in-vivo imaging data of dendritic spines from mouse neocortex [12, 44, 43]. In particular, connections that consist of several contacts are stable for long periods of time (here 150 days) (Fig. 3E inset). The pruned connections are almost exclusively connections that consist of less than four contacts and have relatively small total weight (Fig. 3E). Finally, the lifetime of synaptic connections increases strongly with the total weight of the connection (Fig. 3F), indicating that strong connections are protected against spine turnover by mutual cooperation of contacts.

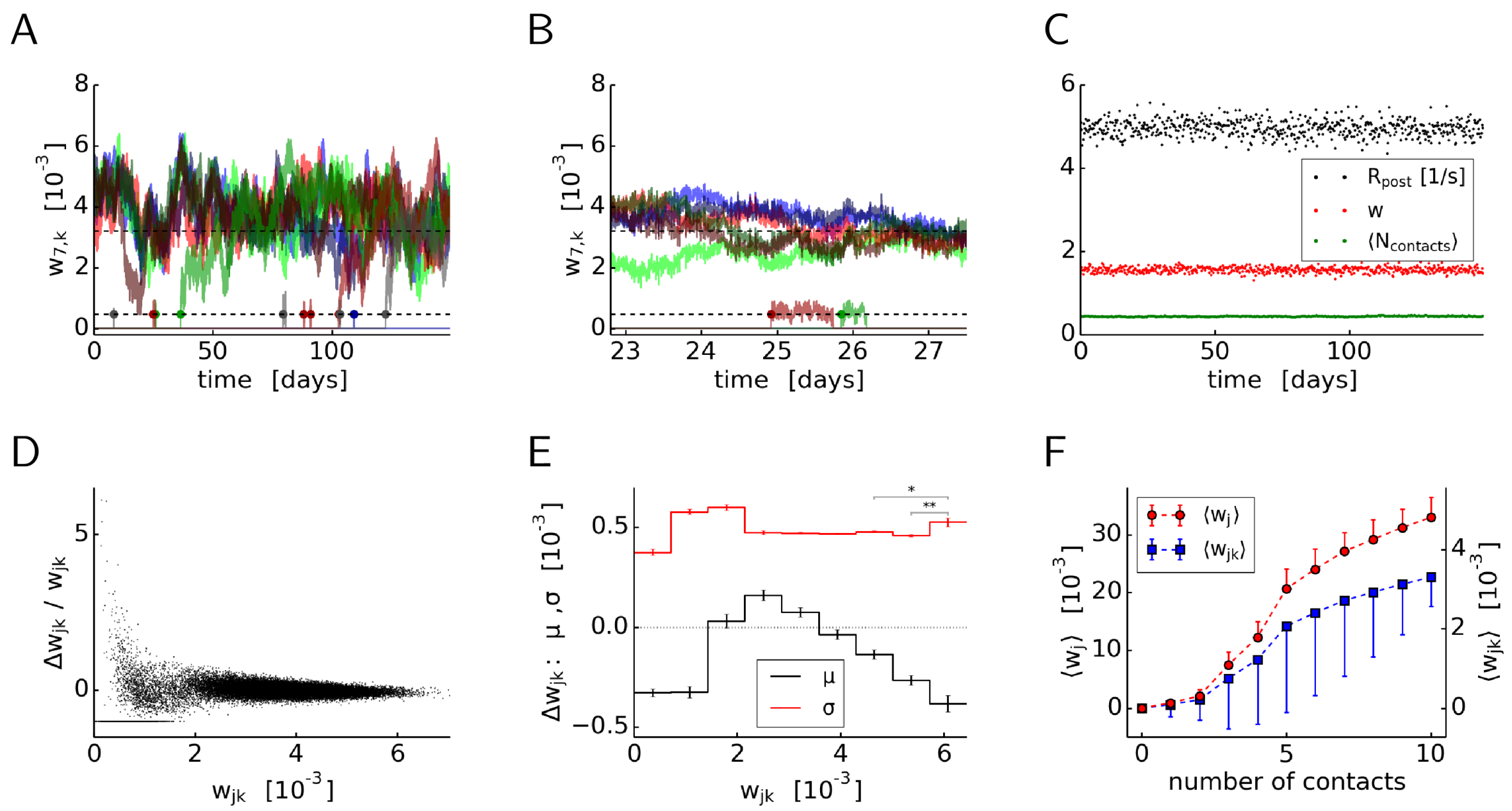

Figure 2: **Dynamics of synaptic contacts in the steady state. A:** Example synaptic connection number $j = 7$, colored lines each correspond to a contact weight $w_{jk}$ over time. New contacts (filled circles) are created with a weight given by the lower dashed line. Long-term stable contacts fluctuate around the upper dashed line, which is the fixed point of $w_*/5$ predicted by theory (see Supplementary Information). **B:** Zoom into A at the time of creation and pruning of two transient synaptic contacts. **C:** Firing rate $R_{\text{post}}$ of postsynaptic neuron (black), total synaptic input $w = \sum_j \sum_k w_{jk}$ summed over all presynaptic neurons and contacts (red, unit-less) and average number of active contacts per synaptic connection (green) over time, each sampled in intervals of 6 hours. **D:** Relative changes of synaptic contact efficacy $\Delta w_{jk}$ within one day, each dot corresponds to the change of one contact in one day, data pooled over all contacts and 150 days of simulation (sampled in intervals of 2 days). The horizontal line of dots at ordinate value $-1$ is due to contacts that were removed within one day. **E:** Weight dependence of contact changes within one day. Mean $\mu$ and standard deviation $\sigma$ of the change of contact weight $\Delta w_{jk}$ within one day, as a function of the contact weight $w_{jk}$. Both $\mu$ and $\sigma$ were estimated from the data shown in D, grouped into $w_{jk}$-intervals as indicated (solid lines). Estimates of $\mu$ and $\sigma$ from less than 5 samples were discarded. Error bars denote the standard error of the mean (SEM). For large contact weights $\sigma$ increases significantly ($*$ indicates $p < 0.1$ in Welsh's two-sided t-test). **F:** Total synaptic connection weight $w_j = \sum_k w_{jk}$ (left axis, red) and contact weight $w_{jk}$ (right axis, blue) as a function of the number of actual contacts ($w_{jk} > 0$), averaged across all synaptic connections. Error bars denote the standard deviation (STD).

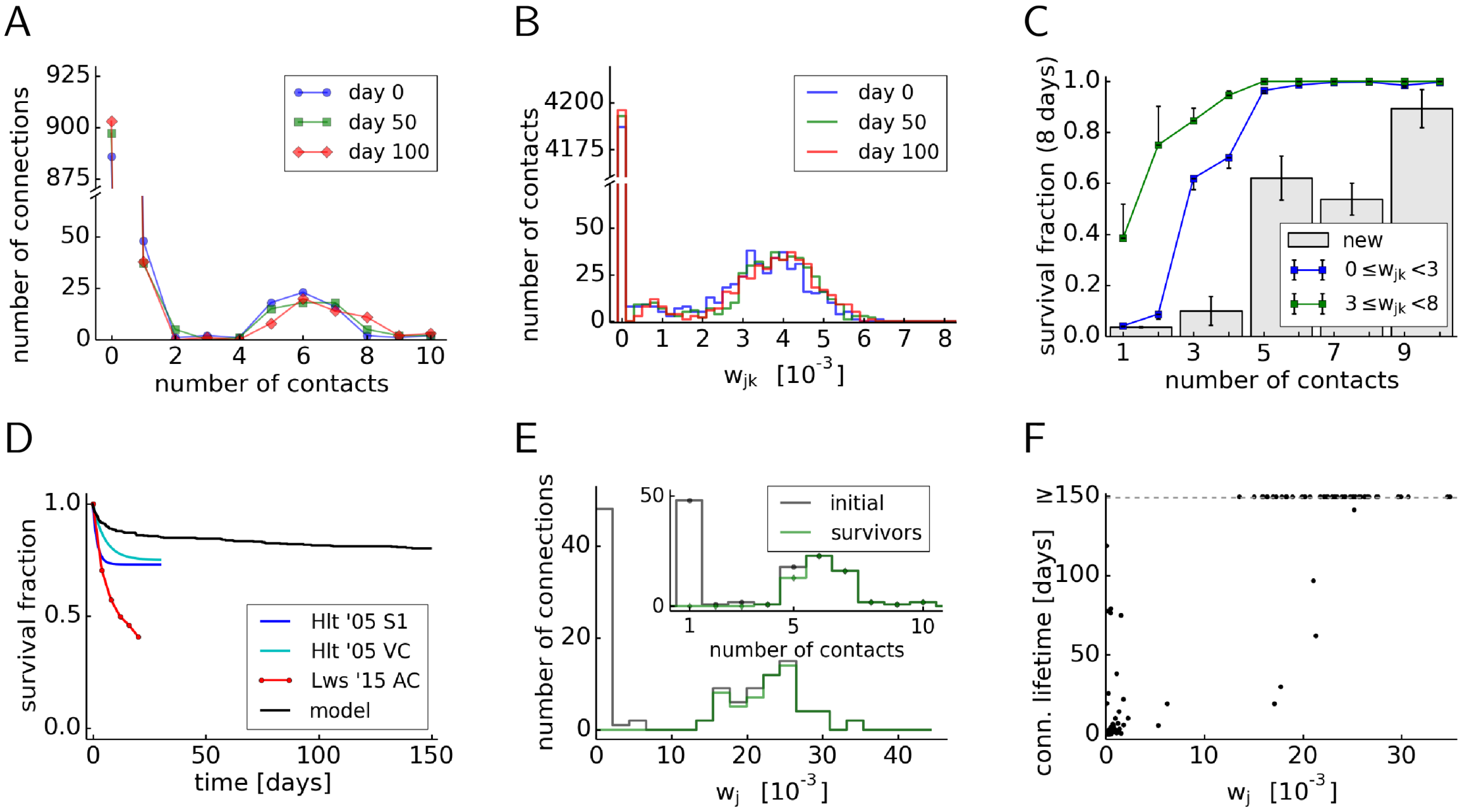

Figure 3: **Statistics of synaptic contacts in the steady state. A:** Histogram of the number of active contacts ($w_{jk} > 0$) of a connection, across all connections $j$. **B:** Histogram of contact weights $w_{jk}$, across all connections $j$ and contacts $k$. The distribution at day 0 is not significantly different from day 100 (two-sample Kolmogorov-Smirnov test (2sKS), p-value 0.910). **C:** Fraction of synaptic contacts that survive for 8 days, as a function of the number of actual contacts in the connection, for newly created contacts (gray bars), and existing contacts of weak or strong weight $w_{jk}$ (colors, see legend) in units of $10^{-3}$. Survival fractions are estimated separately per day and then averaged over days. Estimates from less than 5 samples are discarded; error bars denote SEM. **D:** Fraction of surviving actual synaptic contacts that were present at time $t = 0$ (black: model) in comparison to published experimental dendritic spine survival data in mouse (Hlt '05 S1: somatosensory cortex L5B, age 6 months (exponential decay fit) [12]; Hlt '05 VC: visual cortex L5B, age 3-6 months (exponential decay fit) [12]; Lws '15 AC: auditory cortex [43]). **E**: Histogram of the total weights $w_j$ of the connections present at $t = 0$ (black) and of the surviving connections at day 150 (green). Inset: Histogram of the number of contacts of the connections. The removed connections have small total weight and number of contacts; strong connections with many contacts are rarely removed. Note that new contacts created in the meantime are not considered in this analysis. **F**: Lifetime of entire synaptic connections (dots) during the course of the simulation, as a function of the total connection weight.

## Presynaptic lesions lead to increased formation of persistent contacts

Having established that the steady state of our model has characteristics consistent with experimental data, we investigate whether the model can explain experimentally observed transient dynamics of dendritic spines in response to changes of neuronal input. Experimentally, such input changes may be due to trimming whiskers [42, 16], lesions of the retina [44] or occlusion of one of the eyes [45]. Here we model an abstract lesion experiment in which a fraction of the input neurons suddenly cease to fire (Fig. 4A).

In our lesion model, after ablating 50% of those presynaptic neurons that have active synaptic contacts, we observe a loss of 50% of the synaptic contacts on the postsynaptic neuron after 30 minutes (Fig. 4B). However, this loss is rapidly compensated such that the firing rate and total weight (summed over all presynaptic neurons and all synaptic contacts) are hardly changed throughout the process (Fig. 4B,C). The compensation occurs on two different time scales. First, on the time scale of 10 - 30 minutes, existing synaptic contacts are up-regulated from a pre-lesion value between $(3.3 \pm 1.3) \cdot 10^{-3}$ to a value of $(6.6 \pm 1.3) \cdot 10^{-3}$ measured 30 minutes after the lesion (Fig. 4D). Second, on the slow time scale of 10-30 days after the lesion, the number of presynaptic neurons without postsynaptic contact decreases from 886 pre-lesion to 810 thirty days after lesion while the number of presynaptic neurons with one, two, or three contact points transiently increases (Fig. 4E) suggesting that the plasticity rule 'tests' new connection patterns. Competition between synaptic contact points from different presynaptic neurons and simultaneous cooperation of synaptic contact arising from the same presynaptic neurons leads to pruning or strengthening, so that after 99 days the distribution of contact numbers and synaptic weights is again very similar (but not yet completely identical) to the distributions pre-lesion (Fig. 4E). The gradual recovery of the contact numbers is due to an elevated probability of newly formed contacts to survive and become long-term stable compared to the control condition (Fig. 4G). In a simulated lesion experiment where 20 percent of the active contacts are removed (instead of 50 percent in the simulations so far), 14.6 percent of newly created contacts survive for 8 days or more, consistent with experimental results on dendritic spines in the somatosensory cortex after whisker trimming [16] and significantly above the 7.7 percent of surviving contacts in the control condition (Fig. 4G).

## Synaptic stability and reliability are due to contact multiplicity

To investigate the role of multiple synaptic contacts in our model, we compare its dynamics to the same model restricted to single-contact connections ($n_j = 1$ for all $j$). In the multi-contact model above, with $N = 1000$ presynaptic neurons there are 4633 potential synaptic contacts in total because every presynaptic neuron makes 4.633 potential contacts on average (Fig. 1D, blue line). To maintain the total number of potential synaptic contacts in the system with single-contact connections we therefore increase the number of presynaptic neurons to $N = 4633$, keeping all other parameters the same. Similar to the full model, the single-contact model exhibits steady state dynamics (Fig. 5A) in which postsynaptic firing rate, total weight and synaptic contact number are tightly regulated (data not shown), and the distribution of contact weights is stable over time (Fig. 5B). However, in contrast to the multi-contact model (Fig. 3), in the single-contact model the temporal dynamics of synaptic contact survival do not indicate the presence of a subgroup of stable connections (Fig. 5C), and all connections are prone to random pruning, irrespective of whether they have a small or large synaptic weight (Fig. 5D,E). Thus, multiple synaptic contacts are crucial for the long-term stability of strong synaptic connections, through mutual cooperation of the contacts from the same presynaptic neuron: in multi-contact synaptic connections contacts support each other by exciting the postsynaptic neuron more reliably despite random synaptic transmission failures.

A theoretical analysis further explains the crucial role of contact multiplicity in synaptic transmission. We characterize the fidelity of transmission of a presynaptic spike by the ratio of the (trial-averaged) mean and standard deviation of the evoked postsynaptic potential (PSP), which we call the signal-to-noise ratio (see Methods). In our model, failures of synaptic transmission occur randomly at each synaptic contact, which causes variability of the PSP. If a synaptic connection consists of several contacts, a presynaptic spike is more reliably transmitted than in the case of a single contact (Fig. 5F). If 10 contacts are considered to be the maximum number of active contacts connecting a pair of neurons due to geometrical constraints of the tissue, about 6 contacts give a signal to noise ratio of 80% of the maximum value achievable (irrespective of the synaptic failure rate $p_f$). Note that this number coincides with the typical peak of the model contact numbers in the steady state (cf. Fig. 3A). Indeed, high reliability and low variability of synaptic transmission in layer 5 neurons of the somatosensory cortex of the rat have been observed in glutamate uncaging experiments in acute slices in vitro, and were attributed to the likely presence of multiple synaptic contacts per connection [46].

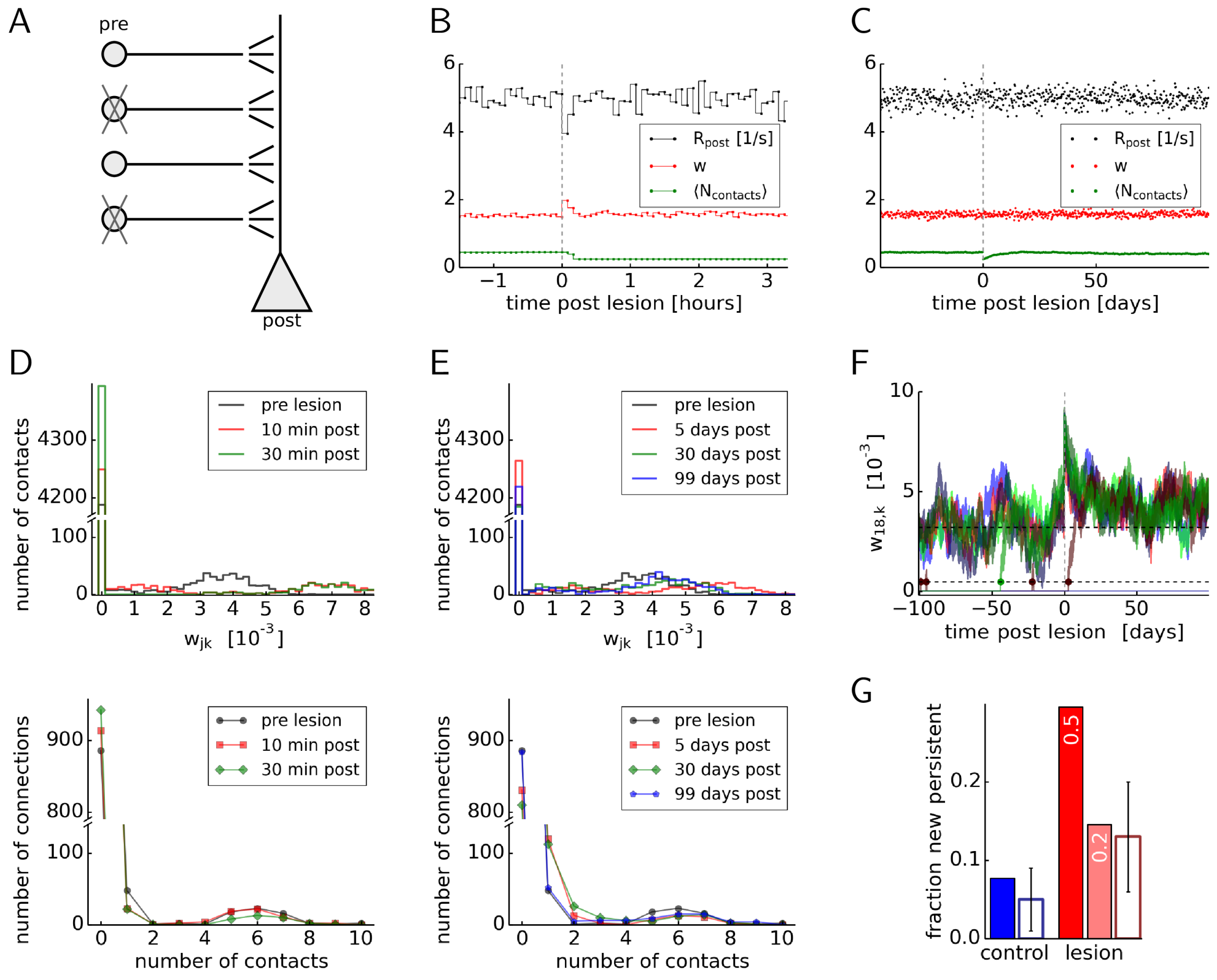

Figure 4: **Rewiring in response to input lesion.** **A:** Schematic of the simulated lesion experiment. A substantial fraction of the presynaptic neurons that have actual synaptic contacts ($w_{jk} > 0$) onto the postsynaptic cell are ablated (set to very low firing rate of 0.1/s; each connected neuron is ablated with probability $p_{\mathrm{lesion}} = 0.5$; unconnected neurons are unaffected) at $t = 0$ (vertical dashes in B-C). **B**-**C:** Firing rate $R_{\mathrm{post}}$ of postsynaptic neuron (black), total synaptic input $w = \sum_{j,k} w_{jk}$ summed over all presynaptic neurons and contacts (red, unit-less) and average number of contacts per synaptic connection (green) over time. (cf. Fig. 2C). After the lesion the average number of contacts (green) quickly drops by about 50% (B), and gradually recovers towards the steady state afterwards (C). **D**-**E:** Histograms of contact weights $w_{jk}$ (top) and of the number of active contacts (bottom), across all connections and contacts, before and after the lesion. Half of the actual synaptic contacts are removed within 30 minutes after the lesion (D, bottom, green), while the remaining half of the contacts potentiates (D, top, green). Within 99 days the distributions gradually recover (E). **F:** Example synaptic connection from presynaptic neuron with index $j = 18$, each colored line corresponds to a contact weight $w_{jk}$ over time, as in Fig. 2A. **G:** Fraction of newly created persistent contacts (which survive for at least 8 days as defined in [16]), control case (blue, cf. Fig. 2) versus lesion model (red). Two lesion models are shown (red bars), marked with their respective value of $p_{\mathrm{lesion}}$. For comparison, data from mouse whisker trimming experiments [16] is shown (open bars, error bars denote STD over observed cells). A smaller lesion with $p_{\mathrm{lesion}} = 0.2$ is in better agreement with the experimental data.

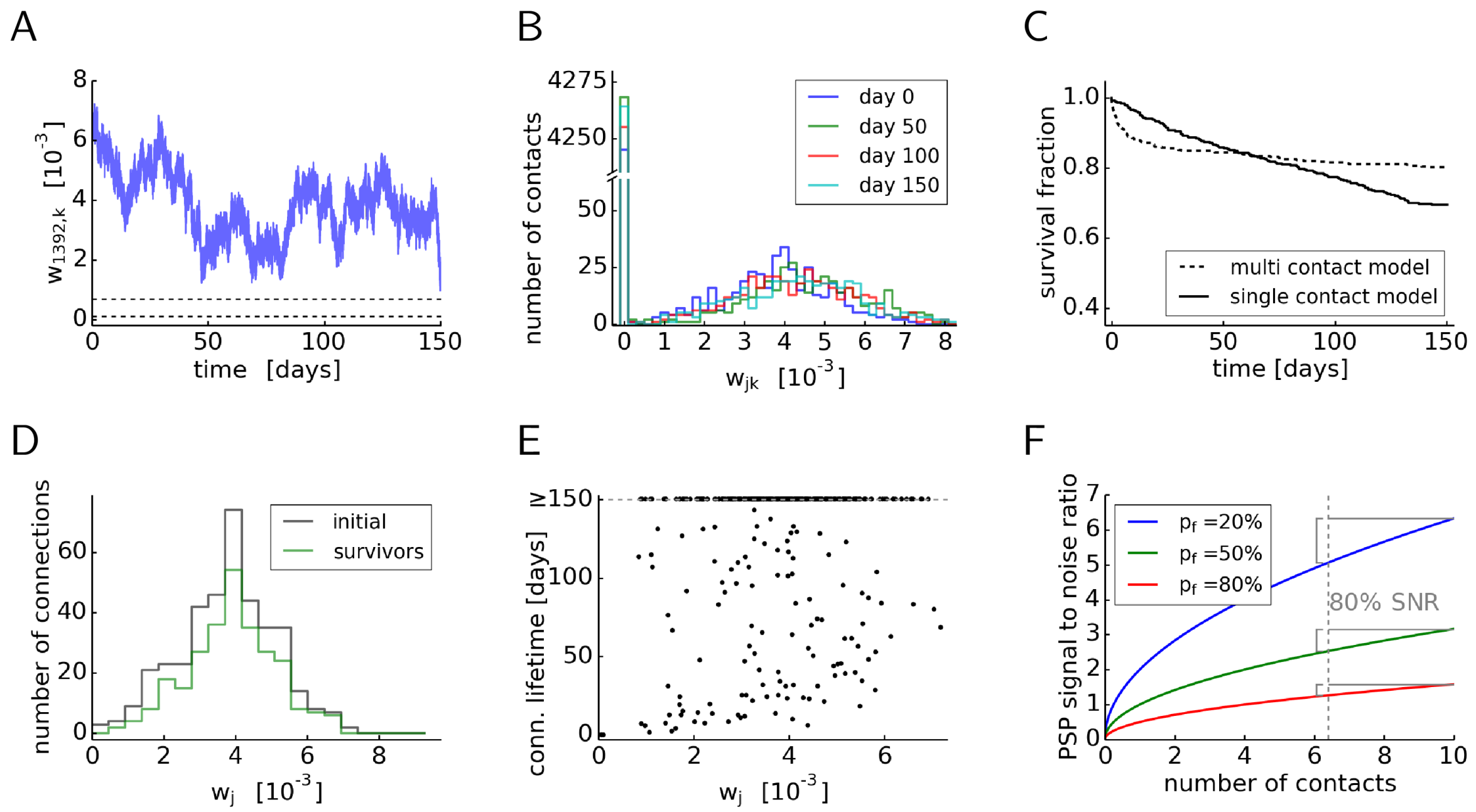

Figure 5: **Single-contact synaptic connections are not long-term stable and have less fidelity. A-E:** Simulation of the plasticity model with only one potential contact per connection ($n_j = 1$ for all $j$). **A:** Example synaptic connection number $j = 1392$, blue line corresponds to contact weight $w_{j1}$ over time, as in Fig. 2A. **B:** Histogram of contact weights $w_{j1}$, across all connections $j$. A steady state is maintained for 100 days, just as in the case of multiple potential contacts (cf. Fig. 3B). **C:** Fraction of surviving actual synaptic contacts (solid line) that were present at time $t = 0$ in comparison to the multi-contact model (dashed line) of Fig. 3D. **D**: Histogram of the weights $w_j = w_{jk}$ of the connections present at $t = 0$ (black) and of the surviving connections at day 100 (green). Connections of small and large weights are removed unspecifically. **E**: Lifetime of entire synaptic connections during the course of the simulation, as a function of the total connection weight. **F:** Theoretical signal to noise ratio (SNR) of the postsynaptic potential (13) in response to a presynaptic spike, in connections with multiple actual contacts and stochastic synaptic failures (probability $p_f$). Dashed line indicates the number of actual contacts for which a SNR of 80% of the maximum is achieved, if 10 contacts is considered to be the maximum.

## Network simulations explain reinnervation of input-deprived cortical barrels

We wondered whether the structural plasticity rule discussed above would allows us to make predictions of structural changes in a large recurrent network of excitatory and inhibitory neurons. We use a network architecture (Fig. 6A) inspired by rodent barrel cortex with three strongly connected cortical populations (representing the columns corresponding to different barrels) of excitatory neurons (exc 1-3), each one preferentially innervated by a thalamic population (tha) that conveys sensory input from one of three whiskers (whi) by strong connections. In the model, connections between different cortical populations and from thalamic populations to non-preferred cortical barrel columns are random and weaker on average (Fig. 6D top left), but all excitatory connections whether strong or weak are subject to the same spike-timing dependent structural plasticity model. A short movement (flick) of the whisker is represented in the model by a small increase in the firing rate of the thalamic neurons which results in turn in a modest increase of the firing rate of the corresponding cortical population (Fig. 6B) riding on top of a spontaneous network activity of about 5Hz (Fig. 6D bottom left). After an initial transient of 7 days of simulated time, we followed the mean connection weights of synapses from tha 3 to exc 3, from tha 2 to exc 3, from exc 3 to exc 3, and from exc 2 to exc 2 during 3 days of simulated time and found no changes (Fig 6C, top), indicating that the average connectivity pattern is globally robust during ongoing activity and random whisker stimulation, despite the fact that the structural plasticity rule is always active.

After 10 days of simulated time, the whisker corresponding to barrel 3 is trimmed. Whisker lesion is modeled by (i) absence of whisker flicks in tha 3 while stimulation of tha 1 and tha 2 continues; and (ii) an exponential decrease of the firing rate of the corresponding thalamic population with a time constant of 5 min to a new baseline level of 0.1Hz. We found that the spiking activity of the network after the lesion remains asynchronous and irregular (Fig. 6B). Within 3 days after the lesion, the recurrent connectivity of the barrel column 3 has increased (Fig. 6C, center) consistent with a recent experiment [47]. The average weight of connections within barrel column 2 is hardly affected by the lesion, but that within barrel column 3 changes substantially. The lateral connections from excitatory neurons of barrel column 2 to 3 and vice versa increase on average. Similarly, the average connection weight of the non-preferred pathway from tha 2 to exc 3 increases whereas the connections in the preferred pathway disappear after the lesion (Fig. 6C, center). We followed the synaptic changes for a total of 50 days after the lesion. Further changes were smaller in magnitude but indicate that the recurrent network slowly continues to reorganize itself into a new connectivity pattern. In particular, the average connectivity from excitatory neurons in barrel column 3 to those in 2 continue to increase.

The increase in the average connection weight of the non-preferred pathway (from tha 2 to exc 3) during the first three days after the lesion could indicate a small increase of all the existing connections, or a stronger increase of a subset of the feed-forward connections. A careful look at the detailed connectivity pattern in Fig. 6D indicates that the latter is the case. Indeed, a subset of neurons located in the excitatory population corresponding to barrel column 3 has become sensitive to stimulation of whisker 2. We first used a clustering algorithm to identify this subset of neurons and then reordered, and color-coded, the excitatory neurons in barrel column 3 according to their responsiveness (yellow shade: newly responsive to whisker 1; red shade: newly responsive to whisker 2; green shade: responsiveness unchanged). The relevance of a reorganization into subsets manifests itself in four different, but consistent ways: (i) the subset of red-shaded neurons in population exc 3 responds more strongly to stimulation of whisker 2 than the average of neurons in exc 3 and nearly as strongly as neurons in the barrel column 2 even though the duration of the response is a bit shorter (Fig. 6E, bottom); (ii) the same subset of red-shaded neurons in population exc 3 has a larger fraction of strong lateral connections to excitatory neurons in barrel column 2 than other neurons in exc 3; (iii) the same subset of red-shaded neurons in population exc 3 receives a larger fraction of strong lateral connections from excitatory neurons in barrel column 2 than from neurons in barrel column 1; (iv) the same subset of red-shaded neurons in population exc 3 receives a larger fraction of strong feed-forward connections from thalamic neurons in group 2 than other neurons in exc 3. Taken together, these four observations suggest that the subset of red-shaded neurons in barrel column 3 has been integrated in the information processing stream of barrel column 2. The same observations can be repeated for the yellow-shaded subgroup of neurons in barrel column 3, except that these neurons have been integrated into barrel column 1. In both cases the integration has been made possible by structural plasticity triggered by the lesion.

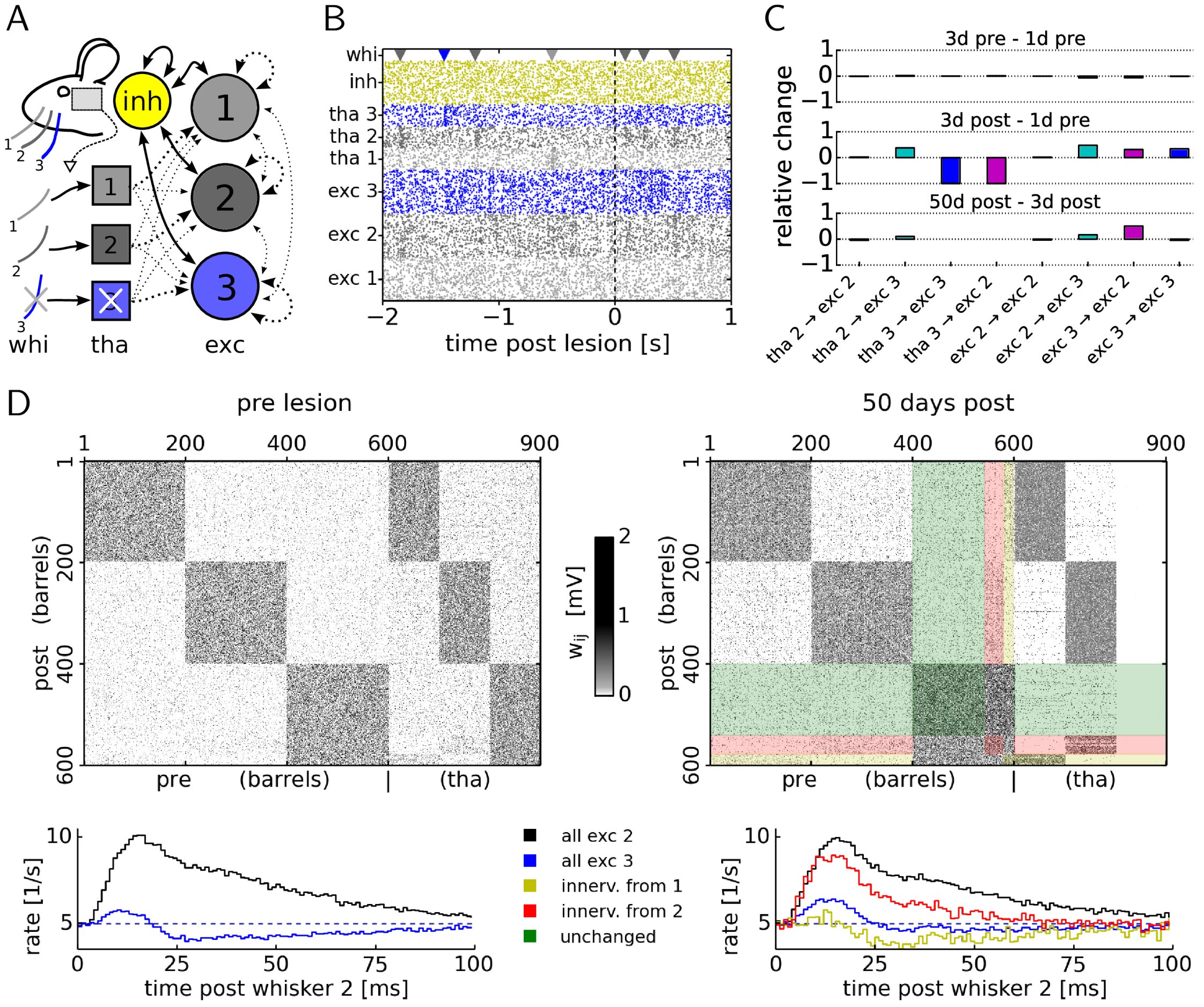

Figure 6: **Structural plasticity in a thalamo-cortical network.** **A:** Schematic of the model. Thalamic neurons (tha) convey sensory input from whiskers (whi) to the recurrent cortical network. Each tha population (squares) projects to one of three cortical "barrels" (circles) of excitatory (exc) neurons. Cortical inhibitory (inh) neurons connect randomly to all barrels. Exc to exc synapses (dotted arrows) are modeled by the structural plasticity model, all other synapses (solid arrows) are static. Plastic exc connections are initialized as follows: Thin arrows $w_{ij} = 0$; thick arrows $w_{ij} = w_*$ (fixed point weight, see Supplementary Information). Inh synapses have a constant weight and no synaptic failures. Tha neurons fire as Poisson processes but increase their firing rate transiently if the corresponding whisker is flicked, which happens randomly with rate 1/s. After 10 days of simulation whisker 3 is trimmed ("lesion"), modeled as a progressive loss of firing of tha 3 neurons (white cross). **B:** Spike raster plot around time of lesion (dashed vertical line), colors as in A. If a whisker is flicked (whi, triangles), the corresponding tha and exc populations respond. **C:** Relative changes of average connection weights $\langle \Delta w_{ij} \rangle / \langle w_{ij} \rangle$ between populations. Changes before (top), around time of lesion (center) and long after (bottom). Strong changes during the first three days post lesion (center) are followed by a slow restructuring process of exc 3 over the following 47 days (bottom). **D (top):** Exc synaptic connection weights (greyscale, $w_{ij}$) just before lesion (left) and 50 days after lesion (right). The comparison of the connection weight matrix before and after lesion shows that loss of tha input to exc 3 caused selective rewiring. Exc 3 neurons (401-600) have been clustered in both graphs according to the inputs they receive at simulation end (see Methods, assignments are indicated by shading). **D (bottom):** Average spike response of exc 2 (black) and exc 3 (blue) neurons in response to whisker 2 flicks, just before trimming of whisker 3 (left) and at simulation end (right) (averaged over 60min of recording). Red and yellow colors denote subgroups of exc 3 identified by clustering. Dashed lines: mean firing rate in recording episode.

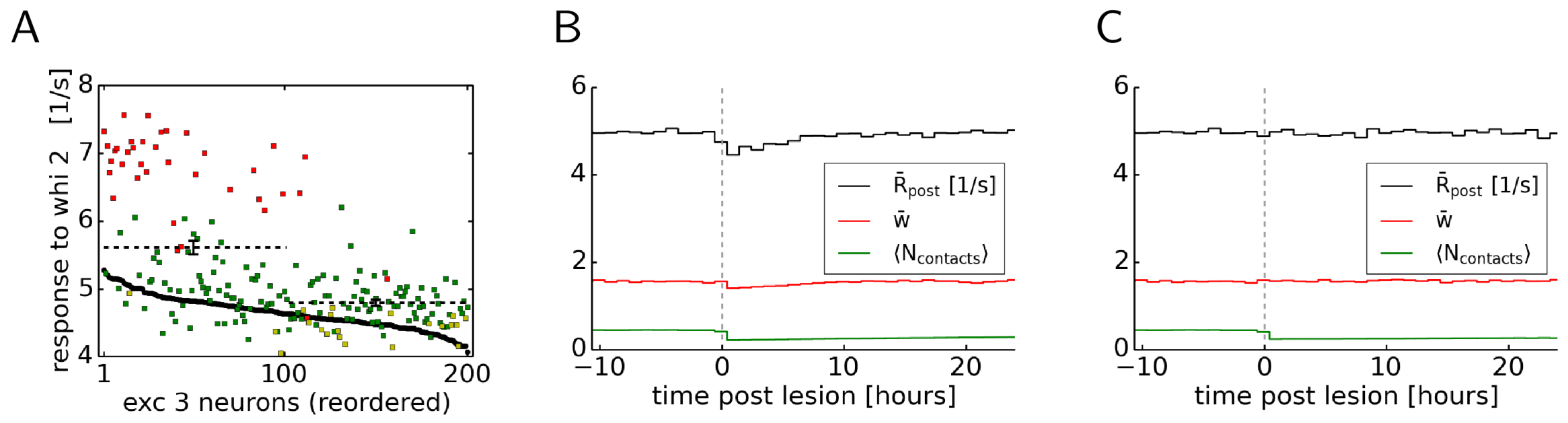

Figure 7: **Restructuring can be predicted from initial responses / effect of bounded weights. A:** Spike rate in response to whisker 2 flicks for each neuron of exc 3, before lesion (black dots) and 50 days post (colored squares), for the thalamo-cortical network simulation shown in Fig. 6. Neurons are ordered according to their pre-lesion response (black). Rates are estimated by counting spikes in a time window of 25ms after each whisker flick. The 100 exc 3 neurons that initially respond strongest to whi 2 are more likely to increase their response to this whisker through structural plasticity (dashed horizontal lines with error bars show mean response 50 days post $\pm$ SEM), and more likely to participate in the cluster that is most strongly innervated by this whisker (red; colors correspond to the cluster assignments of the neurons in Fig. 6D, right). **B:** Rewiring in response to input lesion (as in Fig. 4) with an upper bound of the contact weight so that contact weights cannot grow stronger than twice the fixed point value ($w_{jk} \leq 6.4 \cdot 10^{-3}$). In this case postsynaptic rate and total weight initially decrease in response to the lesion, and recover on a time scale of about 10 hours as new contacts are formed. Simulation data are averaged over consecutive time windows of 1 hour. **C:** For comparison, the data of the lesion simulation without upper bound of Fig. 4B,C is replotted as in (B). Here the remaining contacts quickly (within less than 1 hour) compensate for the loss of input by strongly increasing their weights, so that the postsynaptic rate shows no visible change.

Our simulation results of point (i) above are consistent with experience-dependent receptive field plasticity found experimentally in pyramidal neurons in mouse somatosensory cortex 3-4 days [42] or 20 days [48] after whisker trimming, where neurons that were part of a deprived barrel became responsive to the first-order surrounding whisker, in particular the subset of neurons located in the border region to the neighboring barrel [48]. Observations (ii) - (iv) listed above are predictions of our model. Note that in our simulations, both barrel columns exc 1 and exc 2 are first-order surrounding columns of the deprived column (exc 3) since we have not introduced any distance-dependent connectivity. We emphasize that the parameters of the structural plasticity algorithm are kept fixed throughout the simulation, be it before, during, or after the lesion: First the network connectivity was stationary before the lesion, then it changed significantly during 3 days after the lesion, and finally settled into a new state (Fig. 6C) while structural plasticity has always been 'turned on'. Indeed, individual synaptic contact points continue to grow or disappear even during the phases where the coarse connectivity pattern remains unchanged; cf. Fig. 3.

## Discussion

In this study we linked structural dynamics of synaptic contacts in neuronal networks to spike-timing dependent plasticity (STDP). The implicit coupling between synaptic contacts onto the same postsynaptic neuron through backpropagating action potentials is sufficient to make synaptic contacts from one presynaptic neuron compete with contacts of other presynaptic neurons and cooperate with contacts of the same presynaptic neuron. The resulting high-dimensional non-linear dynamical system has a steady state with properties consistent with those of excitatory synapses in sensory cortices in terms of (i) a bimodal distribution of contact numbers per synaptic connection; (ii) differences in lifetime between strong and weak connections; and (iii) turnover of dendritic spines.

Our model makes at least eight novel predictions.

A) A synaptic contact with a given weight is more stable if it is part of a group of four or five synaptic contacts arising from the same presynaptic neuron than if it is isolated or part of a group of only two synaptic contacts (Fig. 3C).

B) The combination of the known result that strong contacts are more stable than small ones (cf. Fig 2D and [4, 12]) with point A, yields the prediction that connections with larger total synaptic weight (summed over all synaptic contacts) are more stable than small ones (Fig. 3F). This could be measured by correlating the survival time of a synaptic contact with the total EPSP amplitude of the connection.

C) We predict a substantial fraction of synaptic connections that have only one active contact (see Fig. 3A). These synaptic contacts, however, are small and quickly removed (see Fig. 2F). Since weight, PSP amplitude, volume, and size are tightly correlated [5], these weak synapses might escape electrophysiological or visual detection with standard methods which could explain the differences to experimental reports [27, 42], but might be detectable using sub-diffraction resolution imaging [49, 40].

D) After a lesion, the number of connections consisting of exactly two synaptic contacts increases transiently. While this might be expected since the process of building a novel synaptic connection with five contacts has to pass through a transient state with only two contacts, the prediction is that only about a quarter of the two-contact connections actually stabilize to a multi-contact synapse, while the majority disappears again. As a consequence, the number of presynaptic neurons without a connection transiently decreases after a lesion before it increases again to a stable value (Fig. 4D and E).

E) After whisker trimming, the subset of neurons of the deprived cortical column which will be integrated in the signal processing of an adjacent whisker will establish stronger incoming and lateral projections to the cortical column in which they become integrated than other neurons in the deprived column (Fig. 6D).

F) The subset of those neurons of the deprived cortical column that are integrated into a new whisker processing stream do not have to be physical neighbors but can be identified as those who, before the trimming, had already a stronger response to the adjacent whisker (Fig. 7A).

G) The fact that, given a connection, the number of synaptic contacts per pre-post pair in experiments peaks around a finite number (e.g. five contacts) well below the maximal number of potential contacts suggests a novel principle of synaptic plasticity ("learning rule") which normalizes the amount of pre-post correlations. This opens a gateway to a new class of learning rules in unsupervised learning which do not maximize second-order correlations (as done by Oja's rule [50]) but normalize these. While experimentally known homeostasis has focused on the normalization of mean firing rate [51] we suggest that it also is worth while to study a normalization of correlations.

H) Our learning rule assumes that synaptic contact plasticity strongly depends on local traces of correlations that must be processed in the dendritic spines. Therefore, we expect experimental manipulations of biophysical activity traces, such as the calcium concentration, to crucially influence the development of dendritic spines. In line with this hypothesis, effects of local calcium manipulations on dendritic filopodia have indeed been observed [52]. A recent study on in Drosophila larvae further emphasizes the role of presynaptic neurotransmission for the maturation of synaptic terminals [53].

We combined a specific choice of an STDP model with a point neuron model, but several extensions are possible. First, the synaptic plasticity rule used in this paper could also produce branch specific synaptic plasticity and steady-state configurations [54], by choosing a more complex neuron model with non-linear dendritic branches. Branch specificity of the spine dynamics could further be increased by using a voltage-triggered rule [55] instead of a pure STDP paradigm. Second, while the specific choice of STDP rule used in this paper shows, at low frequencies, a symmetric pre-post and post-pre learning window for LTP we can extend our formalism to more realistic STDP models [56, 57]. Third, we chose a learning rule where the evolution of the synaptic contact weights is formally unbounded, but it is straightforward to also include an upper bound. In our model, an explicit upper bound is not necessary (but see end of this section) because synaptic depression limits further growth of contact weights as soon as the total weight of a connection becomes strong, or individual synaptic contacts get large (see Eq. 7 in Methods and Fig. 2E) [58]. For strong weights our learning rule effectively turns from a Hebbian rule to an anti-Hebbian rule, in the sense that a further increase in correlations of pre- and postsynaptic firing leads to a shrinkage of synaptic weight. Similar principles might explain why, depending on experimental preparations, variable STDP rules have been reported in experiments [59]. Fourth, whereas for our recurrent network simulation of Fig. 6 we have used leaky integrate-and-fire neurons, for the mathematical analysis of the system dynamics we chose to describe the activity of the postsynaptic neuron by a Poisson process with a linearly modulated rate. However, because the activity of the postsynaptic model neuron in our simulations fluctuates only in a limited range (see Fig. 2C, black), we expect very similar mathematical results in the case of a non-linear neuron linearized around the operating point.

Previous structural plasticity models of lesion-induced rewiring assumed a homeostasis of the postsynaptic firing rate [37] or spike-timing dependent structural plasticity [35, 38]. Both of these approaches, however, are restricted to structural changes, and do not consider the combined dynamics of continuous weights and discrete structural modifications, which underlie the concurrent up-regulation of weights in response to the lesion in our model (Fig. 4B). Moreover, previous studies did not consider the interplay of synaptic competition between connections, and cooperation within connections, because they either focused on a single presynaptic neuron with multiple contacts [36, 39], multiple presynaptic neurons with single contacts each [35, 38, 60], or non-competitive, purely homeostatic synaptic dynamics [37, 61].

Previous non-structural models of lesion-induced synaptic plasticity highlighted the time scale of homeostasis over hours or days [62]. In our model, lesion-induced changes of synaptic contact weights occur rather quickly. While the average synaptic contact number recovers slowly within about 20 days, within less than 10 minutes after the lesion the contact weights from spared presynaptic neurons are upregulated to compensate for the loss of input (see Figures 4B-D and 6C). In apparent contrast, homeostatic synaptic scaling in response to experimental blocking of postsynaptic firing occurs on the time scale of hours [51] or in response sensory deprivation (likely to correspond to a reduction in presynaptic firing) on the time scale of days [62]. In principle, this discrepancy in the speed of the readjustments of the synaptic weights might be alleviated by low-pass filtering the model's weight changes (1) on a time scale of several hours before applying them to the contacts, but that would make plasticity in general unrealistically slow. Instead we propose to extend the model by a combination of hard bounds [63] and multiplicative interaction of Hebbian and explicit homeostatic processes [62]. With hard bounds set to two times the fixed point weight, the stationary distributions and results of Figs. 1-3 remain unchanged while the recovery of the firing rate after a lesion is slower since individual synaptic contacts cannot grow beyond the hard bound (see Fig 7B). A more complete model could combine the hard bounds with the essential features of the model of Toyoizumi et al [62]. More generally, the interactions of our rule with manifold plasticity mechanisms such as short-term plasticity [23], inhibitory plasticity [64], homeostasis [51, 37, 62], or intrinsic excitability [65, 60] pose interesting challenges for future research.

# Methods

## Spike trains and failures of synaptic transmission.

The activity of the presynaptic neurons $1 \leq j \leq N$ in our model is described by the spike trains $S_j(t) = \sum_m \delta(t - t_j^m)$, where $\{t_j^1, t_j^2, ...\}$ are the spike times of neuron $j$. All presynaptic spike trains are generated by independent Poisson processes with a constant firing rate of $\nu_{\text{pre}} = 5/\text{s}$. Not all presynaptic spikes, however, are transmitted at each synaptic contact that connects neuron $j$ with the postsynaptic neuron. Some spikes are not transmitted due to random transmission failures [66]. Therefore the spike train transmitted at the contact $k$ is given by

$$S_{jk}(t) \quad = \quad \sum_m \delta(t - t_j^m)\, z_{jk}(t_j^m)\,. \tag{2}$$

The spike train $S_{jk}(t) = S_j(t) \cdot z_{jk}(t)$ differs from the presynaptic spike train $S_j$ through multiplication with independent Bernoulli random variables $z_{jk}(t) \in \{0, 1\}$ that describe the stochastic failures of synaptic transmission. In our model, synaptic failures occur randomly and independently with a probability of $p_{\text{f}} = 0.5$, so $z_{jk}(t_j^m)$ is 1 (successful transmission) with probability $1 - p_{\text{f}} = 0.5$. The postsynaptic neuron, in turn, emits the spike train $S_{\text{post}}(t) = \sum_m \delta(t - t^m_{\text{post}})$, where $\{t^1_{\text{post}}, t^2_{\text{post}}, ...\}$ are the spike times. Details on the postsynaptic neuron model are given below.

## Slow and fast traces of pre- and postsynaptic activity.

At each synaptic contact, traces of pre- and postsynaptic activity are formed, which are illustrated in Fig. 1C. The variables $r_{jk}(t)$ and $r_{\text{post}}(t)$ describe low-pass filters of the pre- and postsynaptic activities $S_{jk}(t)$ and $S_{\text{post}}(t)$, defined by the differential equations

$$\tau \frac{d}{dt} r_{jk}(t) = -r_{jk}(t) + S_{jk}(t) \,, \tag{3}$$

$$\tau \frac{d}{dt} r_{\text{post}}(t) = -r_{\text{post}}(t) + S_{\text{post}}(t) \,, \tag{4}$$

with a time constant $\tau = 20\,\text{ms}$, as is typical for an excitatory postsynaptic potential (EPSP) or an STDP window function.

As an estimate of the correlations of pre- and postsynaptic firing on a slower time-scale, each synaptic contact computes $C_{jk}$, as well as a slow trace of the postsynaptic activity $R_{\text{post}}$. These traces are defined by the differential equations

$$\tau_{\text{slow}} \frac{d}{dt} C_{jk}(t) = -C_{jk}(t) + r_{jk}(t) \cdot r_{\text{post}}(t) \tag{5}$$

$$\tau_{\text{slow}} \frac{d}{dt} R_{\text{post}}(t) = -R_{\text{post}}(t) + S_{\text{post}}(t) \tag{6}$$

with a time-constant of $\tau_{\text{slow}} = 1\,\text{min}$. The choice of Eqs. (3), (4) and (5) leads to a symmetric STDP window, but the formalism can be extended to more standard STDP with long-term potentiation and depression by using more traces with different time constants.

## Synaptic contact plasticity.

Synaptic contacts in our model follow a variant of spike-timing dependent plasticity (STDP), see also Fig. 1. Each contact is described by its efficacy (weight) $w_{jk}$, which is the (unit-less) amplitude of the excitatory postsynaptic potential that the contact $k$ evokes upon arrival of an action potential at the presynaptic terminal and in the absence of synaptic failure. Synaptic contacts evolve according to a local STDP rule

$$\frac{d}{dt} w_{jk}(t) = a_2^{\text{corr}} C_{jk}(t) - a_4^{\text{corr}} C_{jk}^2(t) - a_4^{\text{post}} R_{\text{post}}^4(t) - \alpha w_{jk}(t), \tag{7}$$

with the parameters $a_2^{\text{corr}} = 1.94569 \cdot 10^{-6}\,\text{s}$, $a_4^{\text{corr}} = 0.07506 \cdot 10^{-6}\,\text{s}^3$, $a_4^{\text{post}} = 0.02016 \cdot 10^{-6}\,\text{s}^3$ and $\alpha = 2 \cdot 10^{-6}\,\text{s}^{-1}$ (see Supplementary Information S1.2 for details on parameter values). As soon as the dynamics (7) lead to a contact of weight $w_{jk}(t) \leq 0$ its weight is set to 0 and the dynamics cease (inactive contacts, see also Fig. 1B).

Each potential but inactive contact (weight $w_{jk} = 0$) may be created again at random times $t_c$ according to a Poisson process with rate $\lambda_c = 0.019/\text{day}$. In such an event, the weight is set to $w_{jk}(t_c) = w_c$. Further details on the choice of $w_c$ are described below. As suggested by previous works [41, 35] in our model newly created contacts first pass through a "period of grace" of duration $\tau_{\text{gp}} = 15\,\text{min}$, during which the weight is fixed to $w_c$. After the period of grace has passed (for $t \geq t_c + \tau_{\text{gp}}$), the weight dynamics again follow Eq. (7). In the event of synaptic contact creation (at $t_c$) the internal state variables $r_{jk}(t_c)$, $r_{\text{post}}(t_c)$, $C_{jk}(t_c)$ and $R_{\text{post}}(t_c)$ of the contact are each initialized to zero. The period of grace serves as a protected time interval for these variables to equilibrate to the current system state – i.e. to obtain a good estimate of the present pre- and postsynaptic spike rates and correlations.

## Model of postsynaptic activity.

For the simulation of the recurrent network in Fig. 6, the pre- and postsynaptic activity was generated by leaky integrate-and-fire neuron models. However, to perform a mathematical analysis of the model's dynamics, we assume a minimal model of the postsynaptic neuron for the remainder of the study. In the absence of synaptic input from the presynaptic neurons, the postsynaptic neuron fires with a baseline firing rate $\lambda_0 = 1/\text{s}$ (as a Poisson process). We further assume that synaptic inputs cause transient increases of the firing rate on the typical time-scale $\tau$ of an EPSP. We thus model the dynamics of the postsynaptic neuron's firing rate $\lambda(t)$ as

$$\tau \frac{d}{dt}\lambda(t) = -(\lambda(t) - \lambda_0) + \sum_{j=1}^{N} \sum_{k=1}^{n_j} w_{jk}(t) S_{jk}(t - d)\,, \tag{8}$$

where the second term sums all inputs across all $n_j$ synaptic contacts over all $N$ presynaptic neurons, and $d = 1\text{ms}$ denotes the synaptic transmission delay.

## Potential synaptic contacts.

Each presynaptic neuron $j$ may be connected to the postsynaptic neuron by several synaptic contacts, up to a maximum of $n_j$, $1 \leq k \leq n_j$, cf. Fig. 1A. The number $n_j$ of potential contacts of a connection is random, with a probability distribution (Fig. 1D, blue line) estimated in [28] for synapses connecting Layer 5 pyramidal neurons within a maximum distance of $50\,\mu$m in rat barrel cortex. For computational reasons we limited $n$ to a maximum of 10 and renormalized the distribution $P(n)$. Accordingly each value of $1 \leq n \leq 10$ should appear (in expectation) $P(n) \cdot N$ times, where $N$ is the number of presynaptic neurons. In the model we randomly assigned $P(n) \cdot N$ presynaptic neurons to each potential contact number $n$ in order to exactly reproduce $P(n)$, the reported distribution of contact numbers per connection, except for Fig. 5 (single contact case) and Fig. 6 (recurrent network).

## Simulation of the plasticity model.

We performed simulations of the full system using the Neural Simulation Tool (NEST) [67]. We implemented the model as a new multi-contact synapse object, using analytical integration of Eq. (7) as described in Supplementary Information S1.4. Spikes were restricted to a simulation time grid with a step size of $\Delta t = 1\,\text{ms}$. The complete system state ($w_{jk}$, $r_{jk}$, $r_{\text{post}}$, $C_{jk}$, $R_{\text{post}}$, $\lambda$) was recorded in intervals of 5 min. The source code of our plasticity model will be made available on-line as part of NEST (as 'stdp_spl_synapse_hom') upon acceptance of this manuscript for publication; referees can get access by request to the journal editor.

To be able to compare the model dynamics to the adult networks of the referenced experimental studies, we initially simulated the model until a stable synaptic configuration was reached. This steady-state of the system, which is the state at $t = 0$ in Figs. 2 to 4, was obtained by simulating for 100 days after initialization at the theoretically determined fixed point. The initial state was set to $w_{jk} = w_*/5$ for $1 \leq j \leq 100$ and $1 \leq k \leq 5$, and $w_{jk} = 0$ else (see details on the fixed point and definition of $w_*$ below). However, because the numbering of presynaptic neurons in the model is random and arbitrary, it is possible that initially not all presynaptic neurons $1 \leq j \leq 100$ have at least $n_j \geq 5$ potential contacts. Therefore, for any presynaptic neuron $1 \leq j \leq 100$ for which $n_j < 5$, we looked for another neuron $j' > 100$ with $n_{j'} \geq 5$, and exchanged $n_j$ and $n_{j'}$. This procedure allowed us to initialize the system to the theoretical equilibrium state. We also checked that if the system is initialized unconnected ($w_{jk} = 0$ for all $j, k$), fully connected, or randomly connected, the postsynaptic rate, the total weight, and the contact numbers approach a similar steady state (data not shown).

## Expected dynamics of synaptic contacts.

We derive the expected dynamics of the weight of a single synaptic contact under this model. Since dendritic spine plasticity is a slow process compared to the dynamics of action potentials and synaptic transmission, we take the average, denoted as $\langle\cdot\rangle$, of Eq. (7) over realizations of the spike trains ($S$) and synaptic transmission failures ($z$). We obtain

$$\langle \frac{d}{dt} w_{jk}(t) \rangle \approx a_2^{\text{corr}} \langle C_{jk}(t) \rangle - a_4^{\text{corr}} \langle C_{jk}(t) \rangle^2 - a_4^{\text{post}} \langle R_{\text{post}} \rangle^4 - \alpha w_{jk}, \tag{9}$$

which is approximate because squaring and averaging of the terms $C_{jk}$ and $R_{\text{post}}$ have been interchanged. The terms in Eq. (9) can be evaluated as (see Supplementary Information S1.1 for the derivation)

$$\langle R_{\text{post}}\rangle = \langle S_{\text{post}}\rangle = \langle\lambda\rangle = \lambda_0 + \nu_{\text{pre}}(1-p_{\text{f}})\sum_{j=1}^{N}\sum_{k=1}^{n_j} w_{jk} \tag{10}$$

$$\langle C_{jk}\rangle = \langle r_{jk} r_{\text{post}}\rangle = \langle r_{jk}\lambda\rangle = \nu_{\text{pre}}(1-p_{\text{f}})\cdot \left[\frac{1}{2\tau}e^{-d/\tau}(p_{\text{f}} w_{jk} + (1-p_{\text{f}})\sum_{l=1}^{n_j} w_{jl}) + \langle R_{\text{post}}\rangle\right]. \tag{11}$$

Eq. (10) establishes that the postsynaptic rate is determined by the sum of all synaptic weights. Since we assume pre- and postsynaptic neurons to be of the same type, we have chosen the parameters of the plasticity rule such that this rate is on average $\langle R_{\text{post}}\rangle \approx \nu_{\text{post}} = \nu_{\text{pre}}$. Our simulations show that this value is tightly maintained. This implies that the sum of weights $\sum_j \sum_k w_{jk}$ is normalized [68], cf. Eq. (10). Indeed previous theoretical work [58] has shown that terms like $-R_{\text{post}}^4$ in our learning rule lead to a normalization of the total weight.

According to Eq. (11), for small transmission failure probability $p_{\text{f}} \to 0$ the dynamics of the contact $w_{jk}$ (9) is dominated by the total weight $w_j = \sum_k w_{jk}$. For large $p_{\text{f}} \to 1$, the evolution of $w_{jk}$ is independent of $w_j$; so increasing the failure probability $p_{\text{f}}$ gradually decouples the dynamics of the contacts. Inserting Eqs. (10) and (11) into Eq. (9) yields a closed, non-linear system of ordinary differential dynamical equations in $w_{jk}(t)$, with $1 \le k \le n_j$ and $1 \le j \le N$.

In Supplementary Information we describe how the fixed points of the expected contact dynamics can be analyzed and used to calibrate the system (S1.2), and how prototypical trajectories of newly created contacts can be constructed (S1.3).

## Cooperation and competition.

Eqs. (9)-(11) enable us to illustrate the process of cooperation and competition, closely linked to the stabilization of the postsynaptic rate and correlations. The postsynaptic rate does not depend on the weight of any specific synaptic contact, but only on the total input, summed over all weights and contacts; cf. Eq. 10. By contrast, Eq. (11) depends not only on the total input via the rate $R_{\text{post}}$, but in addition also on the individual weight $w_{jk}$ and the total weight $w_j = \sum_l w_{jl}$ arising from the same presynaptic neuron. To study competition, let us consider a uniform state and suppose that all correlations $C_{jk} = c$ and all momentary weights $w_{jk} = w$ for all $j, k$ are small but positive. With $\alpha \ll 1$ in Eq. (9), the dominant evolution is therefore an increase of all weights, driven by the term $a_2^{\text{corr}} c$. However, as the weights increase, the firing rate does so as well and therefore the term $\langle R_{\text{post}}\rangle^4$ eventually stops further growth. This is the essential step of firing rate stabilization. For the same firing rate $\langle R_{\text{post}}\rangle$, some weights will grow further at the expense of others inducing *competition* via the instability of the uniform state, just as in other models [50]. The instability is caused by a positive feedback loop between $\langle dw_{jk}(t)/dt\rangle$ on the left-hand side of Eq. (9) and $w_{jk}$ on the right-hand side of Eq. (11). Going beyond standard plasticity models, Eq. (11) shows that correlations $C_{jk}$ driving the contact weight $w_{jk}$ increase not only proportionally to this specific contact, but also increase with the weight of other contacts $\sum_l w_{jl}$ from the *same* presynaptic neuron. The positive dependence gives rise to *cooperation* between contacts arising from the same neuron. The optimal amount of correlation, and hence cooperation, however, is limited by the term $-a_4^{\text{corr}}\langle C_{jk}\rangle^2$ in Eq. (9). The interplay of Eqs. (9)-(11) therefore stabilizes the firing rate, or total input $\sum_j \sum_k w_{jk}$, as well as the total amount of correlations in active contacts, or total weight $\sum_k w_{jk}$, of those presynaptic neurons that have at least one active contact.

## Contact creation.

With a rate of $\lambda_{\text{c}}$ each potential but inactive contact ($w_{jk} = 0$) may be randomly transformed into an active contact. In such an event, called creation, its weight $w_{jk}$ is set to a low, non-zero value $w_{\text{c}}$, and after the period of grace has elapsed, its dynamics follow Eq. (7) (see Fig. 1B). We have adjusted the value of $\lambda_{\text{c}}$ to fit the experimentally measured turnover ratio of dendritic spines as follows.

The turnover ratio, defined as TOR = (#created + #removed)/(2 · #total · day), was found to be 0.154/day in the adult somatosensory cortex [12], where #created, #removed and #total are the numbers of spines that were created, removed, and the total number observed in one day, respectively. In a steady state we additionally expect that #created = #removed, which implies TOR = #created/(#total · day). Now assume that the model is in a steady-state with a distribution of actual contacts as in Fig. 1D, red line. We derive the creation rate $\lambda_c$ loosely from this distribution, by assuming that approximately 10% of synaptic connections have active contacts, and these have about 5 active contacts. Counting the active synaptic contacts in this case, one would observe about #total $\approx 10\% \cdot N \cdot 5 = 0.5 \cdot N$. On the other hand, the total number of potential contacts in our model is #potential $\approx 4.6 \cdot N$ (Fig. 1D, blue, shows the histogram of $n_j$ across connections). Thus, the number of inactive contacts that are available for creation is #potential − #total. Hence, the expected number of creations is #created = (#potential − #total) · $\lambda_c \cdot 1\,\text{day} = (4.6 - 0.5) \cdot N \cdot \lambda_c \cdot 1\,\text{day}$. Inserting #created and #total into the expression for TOR above, and solving for $\lambda_c$, we obtain $\lambda_c = \text{TOR} \cdot {}^{0.5}/_{4.1} = 0.019/\text{day}$.

## Signal-to-noise ratio of synaptic responses.

To better understand the effects of multiple synaptic contacts in presence of stochastic transmission, let us analyze the postsynaptic response. To this aim, we force the presynaptic neuron $j$ to emit an additional spike at time $t_{\text{pre}}$. For convenience we neglect the synaptic transmission delay here ($d \to 0$), which has no effect on the following reasoning. Assuming constant weights on the short time-scale of synaptic signaling $\tau$, and by averaging Eq. (S2) over all other presynaptic spikes and their synaptic transmission, we obtain the transient postsynaptic response

$$L_j(t \,|\, t_{\text{pre}}) \;=\; \nu_{\text{post}} + \frac{1}{\tau}\theta(t - t_{\text{pre}}) e^{-(t-t_{\text{pre}})/\tau} \sum_{k=1}^{n_j} w_{jk} z_{jk}(t_{\text{pre}}), \tag{12}$$

where we also inserted $\langle R_{\text{post}} \rangle \approx \nu_{\text{post}}$. Note that $L_j(t \,|\, t_{\text{pre}})$ is still a stochastic quantity due to stochastic synaptic transmission $z_{jk}(t_{\text{pre}})$ at the contacts $k$. We may obtain the mean spike-triggered response by averaging over the remaining stochasticity of $z_{jk}$

$$\langle L_j(t \,|\, t_{\text{pre}}) \rangle \;=\; \nu_{\text{post}} + \frac{1}{\tau}\theta(t - t_{\text{pre}}) e^{-(t-t_{\text{pre}})/\tau} (1 - p_{\text{f}}) w_j \,.$$

Thus the average response only depends on the total synaptic weight $w_j$, and not on the configuration of contact weights $w_{jk}$. Similarly, the variance of the response can be derived as

$$\text{var}[L_j(t \,|\, t_{\text{pre}})] \;=\; p_{\text{f}}(1 - p_{\text{f}}) \frac{1}{\tau^2}\theta(t - t_{\text{pre}}) e^{-2(t-t_{\text{pre}})/\tau} \sum_{k=1}^{n_j} w_{jk}^2 \,.$$

Here we see that the contact configuration $w_{jk}$ determines the variance of the postsynaptic response. To further understand these properties, consider a synaptic weight $w_j$ that is made of $n_j$ contacts of weight $w_{jk} = w_j/n_j$. Then the sum of squared weights term in $\text{var}[L_j(t)]$ becomes $\sum_{k=1}^{n_j} w_{jk}^2 = w_j^2/n_j$. For this case we evaluate the signal-to-noise ratio of the synaptic response as

$$\text{SNR}_j \;=\; \frac{\langle L_j(t \,|\, t_{\text{pre}}) \rangle - \nu_{\text{post}}}{\sqrt{\text{var}[L_j(t \,|\, t_{\text{pre}})]}} = \sqrt{\frac{1 - p_{\text{f}}}{p_{\text{f}}} \cdot n_j} \,. \tag{13}$$

Therefore, in presence of synaptic transmission failures, multiple synaptic contacts increase the signal-to-noise ratio of synaptic transmission, proportional to the square root of the number of contacts. Previously, a related result has been found numerically for the mutual information of synaptic inputs and neural outputs via multiple contacts [69].

## Recurrent network model

Here we describe the thalamo-cortical network model presented in Fig. 6. The recurrent cortical network consists of $N_{\text{B}} = 3$ "barrels" of $N_{\text{E}} = 200$ excitatory (exc) neurons each, and $N_{\text{I}} = 200$ inhibitory (inh) neurons that connect without preference to all the barrels (random connections, see details below). All cortical neurons are modeled as leaky integrate-and-fire (LIF) neurons with alpha-function shaped postsynaptic currents ('iaf_psc_alpha' neuron model in NEST simulator). The parameters of the LIF neurons are: membrane time constant $\tau_{\mathrm{LIF}} = 20.6$ms, reset and resting potential $-70$mV, action potential threshold $-55$mV, synaptic time constant 2ms, refractory period 2ms. There are $N_{\mathrm{B}} \cdot N_{\mathrm{E}} + N_{\mathrm{I}} = 800$ cortical neurons in total. All synaptic delays are 1ms.

Each barrel of excitatory neurons is further innervated by thalamic inputs that convey information from the whiskers. There are $N_{\mathrm{T}} = 100$ thalamic input neurons (tha) per barrel. All $N_{\mathrm{B}} \cdot N_{\mathrm{T}} = 300$ tha neurons are modeled as excitatory linear Poisson neurons according to Eq. 8, with a baseline firing rate of $\lambda_0 = 4.5$/s. Each group of thalamic neurons modulates its firing rate in response to flicks of the corresponding whiskers (whi). The sequence of whisker flicks is stochastic and described by Poisson processes $S_1, S_2, S_3$ with rate $\nu_{\mathrm{whi}} = 1$/s. The connection weights from whi to tha are chosen as $w_{\mathrm{whi}} = 0.5$. We assume full connectivity to the respective tha populations, such that each tha neuron responds to each flick of the corresponding whisker. In response to a whisker flick, tha neurons of the receiving population increase their firing rate $\lambda(t)$ transiently by $w_{\mathrm{whi}}/\tau = 25$/s, and subsequently their rate decays back to $\lambda_0$ quickly with time constant $\tau$ (cf. Eq. (8)).

The structural plasticity model (7) describes all exc to exc connections, both thalamo-cortical and intracortical, and is continuously active (except for the first hour after simulation); autapses are excluded. All connections (both from tha to exc and exc to exc) have potential synaptic contacts, but initially the network is connected in a whisker-specific manner as depicted in Fig. 6A, see also below. Because the recurrent network contains many postsynaptic neurons, we need two indices to name a synaptic connection. A contact weight here is denoted by $w_{ijk}$ instead of $w_{jk}$ above, where $i$ denotes the postsynaptic and $j$ denotes the presynaptic neuron, and $k$ the contact. Accordingly, the plasticity rule (7) here reads

$$\frac{d}{dt} w_{ijk}(t) \quad = \quad a_2^{\mathrm{corr}} C_{ijk}(t) - a_4^{\mathrm{corr}} C_{ijk}^2(t) - a_4^{\mathrm{post}} R_i^4(t) - \alpha w_{ijk}(t), \qquad (14)$$

and the total weight of a synaptic connection is $w_{ij}(t) = \sum_{k=1}^{n_{ij}} w_{ijk}(t)$, where $n_{ij}$ is the number of potential contacts for the connection from $j$ to $i$. For simplicity all $n_{ij}$ are drawn from the probability distribution $P(n_{ij})$ (Fig. 1D, blue), irrespective of which group (exc or tha) the neurons $i$ and $j$ belong to.

Because the postsynaptic neurons here are LIF neurons, synaptic efficacies $w_{ijk}$ have to be expressed in units of the PSP (in contrast, above $w_{jk}$ is a unit-less quantity). To match the impulse response function of the LIF neurons receiving an input spike with the fixed point weight $w_*$ to the response of the linear Poisson neurons used above, we scale the synaptic weights as $\hat{w}_{ijk} = \gamma \cdot w_{ijk}$, with $\gamma = 62.82\,\mathrm{mV}$, leading to a typical EPSP amplitude of $\gamma w_* = 1.01\,\mathrm{mV}$. Substituting $\hat{w}_{ijk}$ into Eq. (14) implies that, to maintain the same plasticity dynamics as above, the parameters of the learning rule have to be rescaled according to $a_2^{\mathrm{corr}} \mapsto \gamma a_2^{\mathrm{corr}}$, $a_4^{\mathrm{corr}} \mapsto \gamma a_4^{\mathrm{corr}}$, $a_4^{\mathrm{post}} \mapsto \gamma a_4^{\mathrm{post}}$, $w_0 \mapsto \gamma w_0$. We further inject additional Poisson excitatory and inhibitory input spikes to all LIF neurons, with synaptic weights $\gamma w_*$ (excitation) and $-4\gamma w_*$ (inhibition). Exc neurons receive input rates 1519.2/s (excitation) and 328.3/s (inhibition), inh neurons receive 1391.0/s (excitation) and 351.2/s (inhibition). The scaling factor $\gamma$ and the Poisson process input rates were numerically optimized to match the dynamics of the LIF neuron model to that of the linear Poisson neuron model used above. All other parameters take the same values as before. Note that the network parameters here are chosen such that the system operates approximately at the fixed point of the plasticity dynamics analyzed in Fig. S1.

Apart from its $(N_{\mathrm{B}}[N_{\mathrm{T}} + N_{\mathrm{E}}])(N_{\mathrm{B}}[N_{\mathrm{E}} - 1]) \approx 5.4 \cdot 10^5$ plastic excitatory connections our network also has static synapses. These we set as follows. We choose a connection probability of $p_{\mathrm{conn}} = 1/3$. Each excitatory neuron receives $p_{\mathrm{conn}} N_{\mathrm{I}}$ synapses from randomly chosen inh neurons with a fixed weight of $-(1 - p_{\mathrm{f}})\gamma g w_*$, with $g = 2.5$. Each inhibitory neuron also receives this amount of inhibitory synapses, and $p_{\mathrm{conn}} N_{\mathrm{E}}$ excitatory synapses from randomly chosen exc neurons from each of the $N_{\mathrm{B}}$ cortical barrels, with a fixed weight of $(1 - p_{\mathrm{f}})\gamma w_*$ (these synapses have no transmission failures, therefore the weight is scaled down by the expected transmission rate $(1 - p_{\mathrm{f}})$ of the plastic, stochastic ones).

We initialize the plastic synapses at the theoretically derived fixed point $w_*$, with an expected total of 100 active input connections with 5 contacts each per neuron. So, for each exc-exc and tha-exc connection, we set $w_{ij}(0) = \gamma w_* q_{ij}$ if neuron $i$ and neuron $j$ belong to the same barrel, where $q_{ij}$ is a Bernoulli random number that is 1 with probability $p_{\mathrm{conn}}$ and 0 else (in this way, we get $(N_{\mathrm{T}} + N_{\mathrm{E}})p_{\mathrm{conn}} = 100$ incoming connections per exc neuron in expectation). If $i$ and $j$ are part of different barrels, we set $w_{ij}(0) = 0$. If there are five or more potential contacts ($n_{ij} \geq 5$) in connection $i, j$, we set $w_{ijk}(0) = w_{ij}(0)/5$ for $1 \leq k \leq 5$ and $w_{ijk}(0) = 0$ for $k \geq 5$. If there are less contacts ($n_{ij} < 5$) but $w_{ij}(0) > 0$, we set $w_{ij}(0) = 0$ and look for a connection $i', j'$ that connects the same two groups, has $w_{i'j'}(0) = 0$ and ($n_{i'j'} \geq 5$), and we set $w_{i'j'}(0) = \gamma w_*$ for this connection instead. In Fig. 6D neurons of barrel column exc 3 are ordered according to labels obtained by clustering. We used feature agglomeration based on Ward's hierarchical clustering [70] to assign one of three cluster labels to the vector $\{w_{i1}, w_{i2}, ...\}$ of connection weights (at simulation end) from any exc neuron onto each exc 3 neuron $i$.

Simulations were performed using NEST [67] as described above, except for: (i) contact weights were not recorded, merely the total connection weights $w_{ij}(t)$ and the number of active contacts of each connection; (ii) the system state was recorded only every 60min instead of every 5min above; (iii) we simulated using 44 computing cores in parallel instead of a single core.

## Acknowledgments:

Research was supported by the European Union Seventh Framework Program (FP7) under grant agreement no. 604102 (Human Brain Project, M.D.) and by the Swiss National Science Foundation (200020_147200, A.S.).

# Supplementary Information for "Multi-contact synapses for stable networks: a spike-timing dependent model of dendritic spine plasticity and turnover"

Moritz Deger[1,2], Alexander Seeholzer[1], Wulfram Gerstner[1]

1: School of Computer and Communication Sciences and School of Life Sciences, Brain Mind Institute, École Polytechnique Fédérale de Lausanne, 1015 Lausanne EPFL, Switzerland
2: Institute for Zoology, Faculty of Mathematics and Natural Sciences, University of Cologne, 50674 Cologne, Germany

## S1 Supplementary Methods

### S1.1 Derivation of pre-post correlations.

To derive Eqs. (10) and (11) in the main text, we first write the solution of the differential equations of the traces defined in Eqs. (3) and (4) of the main text,

$$r_\cdot(t) \;=\; \frac{1}{\tau}\int_{-\infty}^{t} e^{-(t-s)/\tau} S_\cdot(s) ds\,, \tag{S1}$$

where $\cdot$ stands for either $jk$ or post. So both $r_{jk}$ and $r_{\text{post}}$ are low-pass filtered versions of presynaptic ($S_{jk}$) and postsynaptic activity ($S_{\text{post}}$), respectively, with time constant $\tau$. Similarly, the expressions $C_{jk}(t)$ and $R_{\text{post}}(t)$ in Eq. (7) of the main text denote low-pass filtered traces (with time-constant $\tau_{\text{slow}}$) of pre-post correlations and the postsynaptic spike train, respectively. Because presynaptic firing and synaptic failures are modeled as independent random variables, taking the expectation value of $r_{jk}$ (S1) with respect to the realizations of $S_{jk}$ yields $\langle r_{jk}\rangle = \langle S_{jk}\rangle = \nu_{\text{pre}}(1-p_{\text{f}})$.

For simplicity, we have assumed that the traces $r_\cdot(t)$ as well as the postsynaptic rate $\lambda(t)$ all evolve on the common time scale $\tau$. The postsynaptic activity $\lambda(t)$ can then, by solving Eq. (8) of the main text and inserting (S1), be written as

$$\begin{aligned}\lambda(t) \;&=\; \lambda_0 + \frac{1}{\tau}\int_{-\infty}^{t} e^{-(t-s)/\tau} \sum_{j=1}^{N}\sum_{k=1}^{n_j} w_{jk}(s) S_{jk}(s-d) ds \\ &=\; \lambda_0 + \sum_{j=1}^{N}\sum_{k=1}^{n_j} w_{jk}(t) r_{jk}(t-d),\end{aligned} \tag{S2}$$

where in the second step we assumed that $w_{jk}$ does not change much on the (fast) time-scale of $\tau$. Since the expectation of $r_{jk}$ is $\nu_{\text{pre}}(1-p_{\text{f}})$ we obtain Eq. (10) of the main text.

Analogously, we compute Eq. (11) of the main text by inserting the explicit solutions of the traces (S1). We arrive at the expression $\langle r_{jk}\lambda\rangle$, which evaluates to

$$\langle r_{jk}\lambda\rangle \;=\; \frac{1}{\tau}\int_{-\infty}^{t} ds\, e^{-(t-s)/\tau}\left[\frac{1}{\tau}\int_{-\infty}^{t} du\, e^{-(t-u)/\tau}\sum_{i,l} w_{il}\langle S_{jk}(s) S_{il}(u-d)\rangle + \langle S_{jk}(s)\rangle\lambda_0\right]. \tag{S3}$$

Here we insert

$$\begin{aligned}\langle S_{jk}(s) S_{il}(u)\rangle \;&=\; \langle S_j(s) z_{jk}(s) S_i(u) z_{il}(u)\rangle = \langle S_j(s) S_i(u)\rangle\langle z_{jk}(s) z_{il}(u)\rangle \\ &=\; C_{ji}(s,u)[\delta_{kl}(1-p_{\text{f}}) + (1-\delta_{kl})(1-p_{\text{f}})^2] + \nu_{\text{pre}}^2(1-p_{\text{f}})^2\,,\end{aligned}$$

where we substituted the covariance function of the spike trains $j$ and $i$, $\mathcal{C}_{ji}(s,u) = \langle S_j(s)S_i(u)\rangle - \langle S_j(s)\rangle\langle S_i(u)\rangle$. For the Poisson processes of our model we have $\mathcal{C}_{ji}(s,u) = \nu_{\text{pre}}\delta_{ij}\delta(s-u)$. Here $\delta_{ij}$ denotes the Kronecker symbol and $\delta(\cdot)$ denotes the Dirac delta function. When summed over $i, l$ together with $w_{il}$, the expression above becomes

$$\begin{aligned}
\sum_{i,l} w_{il}\langle S_{jk}(s)S_{il}(u)\rangle &= \sum_{i,l} w_{il}\mathcal{C}_{ji}(s,u)[\delta_{kl}(1-p_{\text{f}}) + (1-\delta_{kl})(1-p_{\text{f}})^2] + \nu_{\text{pre}}^2(1-p_{\text{f}})^2\sum_i w_i \\
&= (1-p_{\text{f}})\left(\sum_i [p_{\text{f}} w_{ik} + (1-p_{\text{f}})w_i]\mathcal{C}_{ji}(s,u) + \nu_{\text{pre}}[\langle R_{\text{post}}\rangle - \lambda_0]\right) \qquad \text{(S4)} \\
&= \nu_{\text{pre}}(1-p_{\text{f}})\left[\delta(s-u)\left(p_{\text{f}} w_{jk} + (1-p_{\text{f}})w_j\right) + \langle R_{\text{post}}\rangle - \lambda_0\right], \qquad \text{(S5)}
\end{aligned}$$

where we inserted the covariance function $\mathcal{C}_{ji}$ for independent Poisson process inputs in the last step. By insertion of (S5) into (S3) the last equality of Eq. (11) of the main text follows.

Note that our theory can as well describe the more general case in which presynaptic spike trains $S_j$ are non-Poisson, but have stationary covariance function $\mathcal{C}_{ji}(s,u) = \mathcal{C}_{ji}(s-u)$ and stationary mean $\langle S_j(u)\rangle = \nu_{\text{pre}}$. When we insert (S4) instead of (S5) into (S3) we obtain the more general expression for the expected correlation term at the synaptic contact

$$\langle C_{jk}\rangle = \langle r_{jk}\lambda\rangle = (1-p_{\text{f}})\sum_i [p_{\text{f}} w_{ik} + (1-p_{\text{f}})w_i]\,\bar{\mathcal{C}}_{ji} + \nu_{\text{pre}}(1-p_{\text{f}})\langle R_{\text{post}}\rangle\,, \qquad \text{(S6)}$$

with the effective spike train covariance

$$\bar{\mathcal{C}}_{ji} = \tau^{-2}\int_0^\infty\int_0^\infty e^{-(u+s)/\tau}\mathcal{C}_{ji}(u-s+d)\,du\,ds\,.$$

Eq. (S6) shows that the plasticity model is generally sensitive to spike-timing correlations of neurons all across the network.

## S1.2 Mathematical analysis of system dynamics.

As our simulations show, the learning rule maintains a constant postsynaptic firing rate $\nu_{\text{post}}$ by tight regulation of the total input weight. Thus the actual degrees of freedom of the multi-connection multi-contact system reside in the configuration of which of the connections $w_j$ are strong or weak, and through which number of contacts $w_{jk}$ this strength is achieved. To understand these dynamics, we consider the expected change of the weight of a single contact $d\langle w_{jk}\rangle/dt$ assuming the total input $w = \sum_{j=1}^N w_j$ is constant and given by Eq. (10) in the main text. This analysis is displayed in Fig. S1.

With the parameters of the learning rule (Eq. (7) of the main text) our mathematical analysis shows that the steady state has the following properties: (a) There is a stable fixed point of Eq. (9) (marked by the circle in Fig. S1), such that if only 10% of the synaptic connections are active (consistent with [1, 2]), with total weight $w_j = w_*$ each, the postsynaptic firing rate of $\nu_{\text{post}} = \nu_{\text{pre}} = 5/\text{s}$ is achieved. From Eq. (10) in the main text follows that this fixed point weight has to be $w_* = (\nu_{\text{post}} - \lambda_0)/[(1-p_{\text{f}})\nu_{\text{pre}}\cdot 10\%\cdot N] = 0.016$. The creation weight $w_{\text{c}}$ (cross in Fig. S1) was set arbitrarily to 15% of the fixed point weight of a contact, $w_{\text{c}} = 0.15\, w_*/5$ (our results are insensitive against the exact value). (b): Contacts only have a stable fixed point of $w_{jk}$ in connections that contain at least three active contacts (dash-dotted lines in Fig. S1A). In connections with less than three active contacts (dashed lines in Fig. S1A), $d\langle w_{jk}\rangle/dt$ is generally smaller than 0, so that these contacts are removed eventually. If (b) is fulfilled, contacts in connections with about 5 contacts, as observed experimentally in [1], are stable.

To calibrate our plasticity model (Eq. (7) of the main text) we used a numerical optimization procedure to find parameters $a_2^{\text{corr}}$, $a_4^{\text{corr}}$, $a_4^{\text{post}}$ and $\alpha$ that fulfill these constraints. As the configuration of fixed points of Eq. 9 of the main text is insensitive to a common rescaling of these parameters, we manually adjusted the scale of the parameters to achieve a good agreement of the temporal dynamics of the contacts (cf. Fig. 2A,D,E in the main text) with the experimental references.

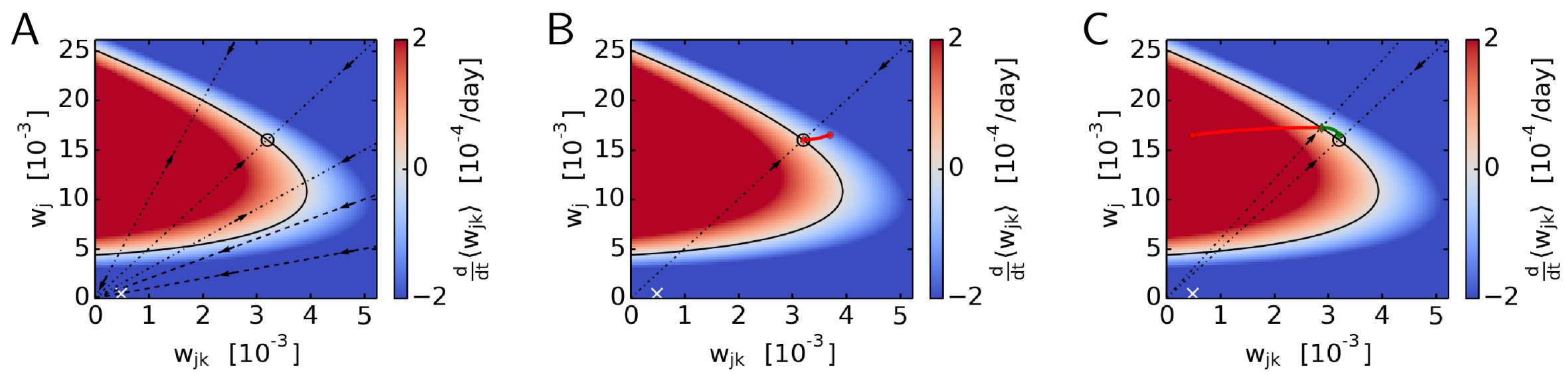

Figure S1: **Expected dynamics of synaptic contacts.** **A:** Positive or negative change (red or blue color, Eq. 9) of a synaptic contact as a function of its weight $w_{jk}$ (horizontal axis) and the total weight of the connection from presynaptic neuron $j$ ($w_j$, vertical axis), under the assumption that the sum of all weights of all presynaptic neurons is implicitly normalized via Eq. (10). Straight lines mark synapses where $w_j$ is a multiple of $w_{jk}$, i.e. synapses consisting of 1 and 2 contacts (dashed), or 3, 5 and 10 contacts (dash-dotted). The circle marks the stable fixed point ($w_*/5$, $w_*$) of the dynamics (for 5 actual contacts), the white cross marks the point of a newly created contact in a previously inactive connection ($w_c$, $w_c$). The black curve marks combinations of $w_{jk}$ and $w_j$ that have zero expected change ($w_{jk}$ nullcline). **B**-**C**: Expected trajectories after a perturbation. We assume that in a synaptic connection $j$ with $n_j$ contacts, $n_j - 1$ contacts have identical values while a single contact $w_{jk}$ is perturbed. **B:** In a synaptic connection with five contacts (fixed point at $w_j = w_*$, circle), four contacts have the same weight $w_*/5$ while the contact $w_{jk}$ is perturbed by a small amount $w_\Delta$. Its trajectory (red line) starts at ($w_*/5 + w_\Delta$, $w_* + w_\Delta$) and evolves back towards the fixed point (circle). **C:** Five contacts have existed for a long time and each take a value $w_{jl} = w_*/5$ consistent with the fixed point (circle), when spontaneously a new contact $w_{jk}$ is created at its creation value $w_c$. The trajectory of the new contact (red line) starts at ($w_c$, $w_* + w_c$) and evolves towards the new fixed point for 6 contacts. The previously existing contacts also evolve towards the new fixed point (green line).

## S1.3 Expected trajectories in response to perturbations.

The expected dynamics (Eq. (9) in the main text) allow us to predict trajectories of ($w_{jk}, w_j$). In Fig. S1B we show that, in expectation, if a contact weight $w_{jk}$ in an active connection at steady-state is perturbed, the perturbation decays. In Fig. S1C, we display what happens when a new contact is created at the steady state in an active connection with five contacts. We find that the new contact as well as the five existing contacts move to a new fixed point corresponding to six stable contacts. As we have seen above, a synaptic connetion consisting of three or more contacts is stable in expectation. In Fig. S2 we consider the creation of synaptic contacts in an inactive connection, in which no other active contacts exist. Both for a single newly created contact (Fig. S2A) and for two simultaneously created new contacts (Fig. S2B), the new contacts are expected to approach zero. A phase plane analysis of the dynamics of two contacts (Fig. S2C) further confirms that, in expectation, all connections of only two contacts are eventually removed because of the absence of stable fixed points. However, because of a region on the phase plane where the dynamics are slow, a connection with two contacts has a lifetime that is sufficiently long so that occasionally a third contact may be added. Still, due to the low creation rate we expect this event to be rare. These theoretical considerations suggest that the system is indeed calibrated to have a steady state that is qualitatively consistent with the experimental contact number distributions, in the sense that stable connections have at least three synaptic contacts.

## S1.4 Analytical integration of synaptic dynamics.

As we show here, the dynamics of the plasticity model can be integrated analytically in the absence of spikes. This allows us to iterate the dynamics from spike to spike precisely and efficiently, and to implement it as an event-driven algorithm in NEST [3].

If for all $t \in (t_0, t_1)$ there are no transmitted pre- nor postsynaptic spikes, $S_{jk}(t) = S_{\text{post}}(t) = 0$, then the traces defined in Eqs. (3), (4) and (6) in the main text decay exponentially as $r_{jk}(t) = r_{jk}(t_0)\exp(-(t - t_0)/\tau)$, $r_{\text{post}}(t) = r_{\text{post}}(t_0)\exp(-(t - t_0)/\tau)$ and $R_{\text{post}}(t) = R_{\text{post}}(t_0)\exp(-(t - t_0)/\tau_{\text{slow}})$, respectively.

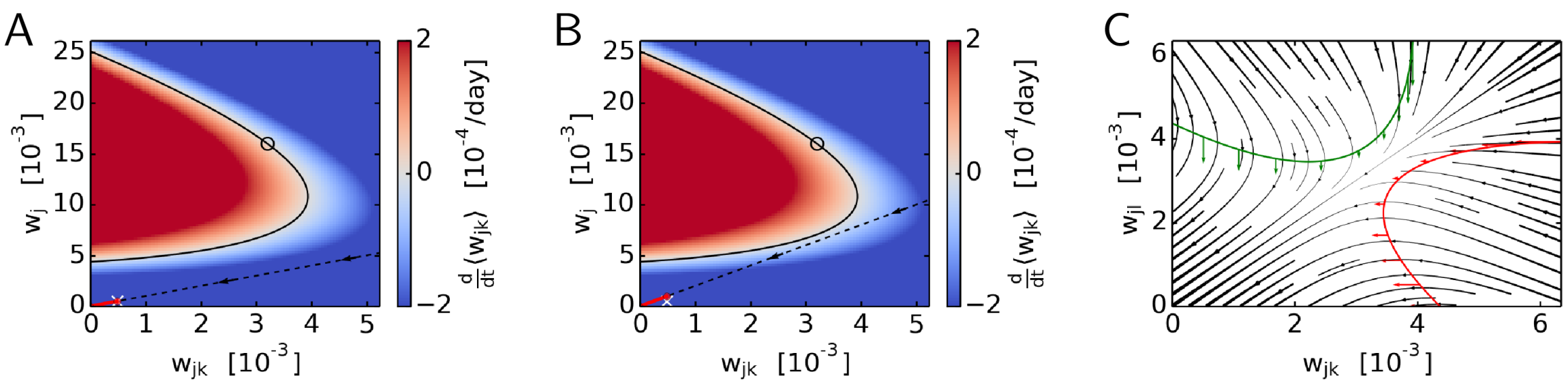

Figure S2: **Transient synaptic contacts in inactive connections. A:** New contact in a connection without other actual contacts. The new contact approaches $w_{jk} = 0$ and is removed (red line). **B:** Simultaneous creation of two new contacts in a connection without other contacts. Both of the contacts identically approach $w_{jk} = 0$ and are removed (red line). **C:** Phase plane analysis of the two contacts weights $w_{jk}$ and $w_{jl}$ for a connection made of 2 contacts ($k \neq l$). The nullclines of $w_{jk}$ (green) and $w_{jl}$ (red) do not intersect, therefore the system does not have a fixed point, but there is a region where the nullclines are close to each other where the flow is rather slow. Colored arrows denotes the gradient $^{d}/_{dt}(w_{jk}, w_{jl})$ on the nullclines (arrow length corresponds to gradient magnitude in arbitrary units). Black streamlines indicate direction of flow of the dynamics, thickness of arrows denotes speed. Visibly, all trajectories converge to $(0, 0)$, yet the ghosts of an unstable fixed point in the region where nullclines are close slows the dynamics sufficiently to allow for the occasional creation of additional contacts.

Inserting these functions into Eq. (5) of the main text and solving the differential equation yields

$$C_{jk}(t) \;=\; C_{jk}(t_0)e^{-(t-t_0)/\tau_{\text{slow}}} + r_{jk}(t_0)r_{\text{post}}(t_0)\frac{e^{-2(t-t_0)/\tau} - e^{-(t-t_0)/\tau_{\text{slow}}}}{1-2\tau_{\text{slow}}/\tau}\,. \tag{S7}$$

Similarly also $w_{jk}(t)$ can be solved for analytically. We insert the solution for $R_{\text{post}}(t)$ and Eq. (S7) and solve Eq. (7) of the main text. The resulting analytical expression for $w_{jk}(t)$ is a generalized Dirichlet polynomial given by

$$w_{jk}(t) \;=\; \frac{1}{c}\sum_{i=1}^{7} a_i e^{b_i(t-t_0)}\,, \tag{S8}$$

with the parameters

$$\begin{aligned} a_1 \;=\;& 2a_4^{\text{corr}} r_{jk}(t_0)r_{\text{post}}(t_0)\tau^2(-4+\alpha\tau)(-2+\alpha\tau)\tau_{\text{slow}}(-(C_{jk,\text{post}}(t_0)\tau) + \\ & r_{jk}(t_0)r_{\text{post}}(t_0)\tau + 2C_{jk,\text{post}}(t_0)\tau_{\text{slow}})(-4+\alpha\tau_{\text{slow}})(-2+\alpha\tau_{\text{slow}})(-1+\alpha\tau_{\text{slow}})\,, \end{aligned}$$

$$\begin{aligned} a_2 \;=\;& a_2^{\text{corr}}(-4+\alpha\tau)(-2+\alpha\tau)(-(r_{jk}(t_0)r_{\text{post}}(t_0)\tau) + \\ & C_{jk,\text{post}}(t_0)(\tau-2\tau_{\text{slow}}))(\tau-2\tau_{\text{slow}})\tau_{\text{slow}}(-4+\alpha\tau_{\text{slow}})\cdot \\ & (-2+\alpha\tau_{\text{slow}})(-2\tau_{\text{slow}}+\tau(-1+\alpha\tau_{\text{slow}}))\,, \end{aligned}$$

$$\begin{aligned} a_3 \;=\;& -(a_4^{\text{corr}}(-4+\alpha\tau)(-2+\alpha\tau)\tau_{\text{slow}}(-(C_{jk,\text{post}}(t_0)\tau) + r_{jk}(t_0)r_{\text{post}}(t_0)\tau + \\ & 2C_{jk,\text{post}}(t_0)\tau_{\text{slow}})(-4+\alpha\tau_{\text{slow}})(-1+\alpha\tau_{\text{slow}})(-2\tau_{\text{slow}}+\tau(-1+\alpha\tau_{\text{slow}})))\,, \end{aligned}$$

$$\begin{aligned} a_4 \;=\;& -(a_4^{\text{post}}R_{\text{post}}^4(t_0)(-4+\alpha\tau)(-2+\alpha\tau)(\tau-2\tau_{\text{slow}})^2\tau_{\text{slow}}\cdot \\ & (-2+\alpha\tau_{\text{slow}})(-1+\alpha\tau_{\text{slow}})(-2\tau_{\text{slow}}+\tau(-1+\alpha\tau_{\text{slow}})))\,, \end{aligned}$$

$$\begin{aligned} a_5 \;=\;& -(a_4^{\text{corr}} r_{jk}^2(t_0)r_{\text{post}}^2(t_0)\tau^3(-2+\alpha\tau)(-4+\alpha\tau_{\text{slow}})\cdot \\ & (-2+\alpha\tau_{\text{slow}})(-1+\alpha\tau_{\text{slow}})(-2\tau_{\text{slow}}+\tau(-1+\alpha\tau_{\text{slow}})))\,, \end{aligned}$$

$$a_6 = a_2^{\mathrm{corr}} r_{jk}(t_0) r_{\mathrm{post}}(t_0) \tau^2 (-4+\alpha\tau)(\tau - 2\tau_{\mathrm{slow}})(-4+\alpha\tau_{\mathrm{slow}}) \cdot (-2+\alpha\tau_{\mathrm{slow}})(-1+\alpha\tau_{\mathrm{slow}})(-2\tau_{\mathrm{slow}} + \tau(-1+\alpha\tau_{\mathrm{slow}}))\,,$$

$$\begin{aligned} a_7 = {} & (\tau - 2\tau_{\mathrm{slow}})^2 (w_{jk}(t_0)(-4+\alpha\tau)(-2+\alpha\tau)(-4+\alpha\tau_{\mathrm{slow}})(-2+\alpha\tau_{\mathrm{slow}}) \cdot \\ & (-1+\alpha\tau_{\mathrm{slow}})(-2\tau_{\mathrm{slow}} + \tau(-1+\alpha\tau_{\mathrm{slow}})) + a_2^{\mathrm{corr}}(-4+\alpha\tau)(-4+\alpha\tau_{\mathrm{slow}}) \cdot \\ & (-2+\alpha\tau_{\mathrm{slow}})(r_{jk}(t_0) r_{\mathrm{post}}(t_0)\tau + C_{jk,\mathrm{post}}(t_0)(2-\alpha\tau)\tau_{\mathrm{slow}}) \cdot \\ & (-2\tau_{\mathrm{slow}} + \tau(-1+\alpha\tau_{\mathrm{slow}})) + (-2+\alpha\tau)(-1+\alpha\tau_{\mathrm{slow}})(a_4^{\mathrm{post}} R_{\mathrm{post}}^4(t_0) \cdot \\ & (-4+\alpha\tau)\tau_{\mathrm{slow}}(-2+\alpha\tau_{\mathrm{slow}})(-\tau - 2\tau_{\mathrm{slow}} + \alpha\tau\tau_{\mathrm{slow}}) + a_4^{\mathrm{corr}}(-4+\alpha\tau_{\mathrm{slow}}) \cdot \\ & (2r_{jk}^2(t_0) r_{\mathrm{post}}^2(t_0)\tau^2 - C_{jk,\mathrm{post}}(t_0)(C_{jk,\mathrm{post}}(t_0) + \\ & 2r_{jk}(t_0) r_{\mathrm{post}}(t_0))\tau(-4+\alpha\tau)\tau_{\mathrm{slow}} + C_{jk,\mathrm{post}}^2(t_0)(-4+\alpha\tau)(-2+\alpha\tau)\tau_{\mathrm{slow}}^2)))\,, \end{aligned}$$

$b_1 = -(1/\tau_{\mathrm{slow}} + 2/\tau)$, $b_2 = -1/\tau_{\mathrm{slow}}$, $b_3 = -2/\tau_{\mathrm{slow}}$, $b_4 = -4/\tau_{\mathrm{slow}}$, $b_5 = -4/\tau$, $b_6 = -2/\tau$, $b_7 = -\alpha$, and

$$c = (-4+\alpha\tau)(-2+\alpha\tau)(\tau - 2\tau_{\mathrm{slow}})^2(-4+\alpha\tau_{\mathrm{slow}}) \cdot (-2+\alpha\tau_{\mathrm{slow}})(-1+\alpha\tau_{\mathrm{slow}})(-2\tau_{\mathrm{slow}} + \tau(-1+\alpha\tau_{\mathrm{slow}}))\,.$$

We can thus integrate the dynamics of the plasticity model exactly between any two spikes by evaluating Eq. (S8).

At any time $t$ at which there is a spike ($S_{jk}(t) \neq 0$ or $S_{\mathrm{post}}(t) \neq 0$), we merely have to increment the traces: $r_{jk}(t) \leftarrow r_{jk}(t) + 1/\tau$, $r_{\mathrm{post}}(t) \leftarrow r_{\mathrm{post}}(t) + 1/\tau$, or $R_{\mathrm{post}}(t) \leftarrow R_{\mathrm{post}}(t) + 1/\tau_{\mathrm{slow}}$, respectively. Then the solutions of the differential equations for the subsequent inter-spike-interval are as given above, with $t_0 \leftarrow t$.

When iterating the time-evolution of $w_{jk}$ from spike to spike by Eq. (S8), however, we need to make sure that $w_{jk}(t)$ does not cross the zero line at any intermediate $t \in (t_0, t_1)$ (zero-crossings are important because they cause pruning of the contact). Therefore we apply Theorem 4.7 of [4], which provides a criterion to rule out zero-crossings of $w_{jk}(t)$ for any $t > t_0$. If zero crossings cannot be ruled out by this criterion, we evaluate Eq. (S8) for all $t$ in the interval $(t_0, t_1)$, on the simulation time grid $\Delta t$, to guarantee that no zero-crossings of $w_{jk}$ were missed.